\documentclass[12pt]{iopart} 
\usepackage[utf8]{inputenc}
\usepackage[english]{babel}
\usepackage{footmisc}
\expandafter\let\csname equation*\endcsname\relax
\expandafter\let\csname endequation*\endcsname\relax
\usepackage{amsmath}
\usepackage{cases}
\usepackage{graphicx}
\usepackage{upgreek} 
\usepackage{stmaryrd}
\usepackage{caption}
\usepackage{subcaption}
\usepackage{hhline}
\usepackage{url}
\usepackage{cite}
\usepackage{hyperref}

\usepackage{ulem} 

\usepackage{todonotes}

\graphicspath{ {./}}

\newcommand{\CH}[2]{\text{CH}\ensuremath{_{#1}^{#2}}}

\renewcommand{\H}[2]{\text{H}\ensuremath{_{#1}^{#2}}}

\begin{document}

\title[Neutral dissociation of methane by electron impact $\ldots$]{Neutral dissociation of methane by electron impact and a complete and consistent cross section set}
\author{Dennis Bouwman$^1$, Andy Martinez$^{1,2}$, Bastiaan J. Braams$^1$, Ute Ebert$^{1,2}$}
\address{$^1$ Centrum Wiskunde \& Informatica (CWI), Amsterdam, The Netherlands}
\address{$^2$ Department of Applied Physics, Eindhoven University of Technology, PO Box 513, 5600 MB Eindhoven, The Netherlands}

\eads{\mailto{Dennis.Bouwman@cwi.nl}}

\begin{abstract}
\noindent
We present cross sections for the neutral dissociation of methane, in a large part obtained through analytical approximations. With these cross sections the work of Song {\it et al.} [J.\ Phys.\ Chem.\ Ref.\ Data, \textbf{44}, 023101, (2015)] can be extended which results in a complete and consistent set of cross sections for the collision of electrons with up to 100~eV energy with methane molecules. Notably, the resulting cross section set does not require any data fitting to produce bulk swarm parameters that match with experiments. Therefore consistency can be considered an inherent trait of the set, since swarm parameters are used exclusively for validation of the cross sections. Neutral dissociation of methane is essential to include (1) because it is a crucial electron energy sink in methane plasma, and (2) because it largely contributes to the production of hydrogen radicals that can be vital for plasma-chemical processes. Finally, we compare the production rates of hydrogen species for a swarm-fitted data set with ours. The two consistent cross section sets predict different production rates, with differences of $45\%$ (at $100$\,Td) and $125\%$ (at $50$\,Td) for production of \H{2}{} and a similar trend for production of H. With this comparison we underline that the swarm-fitting procedure, used to ensure consistency of the electron swarm parameters, can possibly deteriorate the accuracy with which chemical production rates are estimated. This is of particular importance for applications with an emphasis on plasma-chemical activation of the gas.
\end{abstract}


\submitto{\PSST on June 4, 2021}
\maketitle
\ioptwocol 

\section{Introduction}
\subsection{Methane-containing plasmas}
There are many types of plasma that contain methane (\CH{4}{}). Proper models of their properties require cross sections for the collisions of electrons with methane molecules. The present study was particularly motivated by applications such as plasma-assisted combustion of air-methane mixtures, where electron impact dissociation accounts for most of the plasma-produced radicals during the discharge phase \cite{starikovskaia_plasma_2006, starikovskiy_plasma-assisted_2013}. Another combustion-related application is the production of hydrogen fuel through electron impact dissociation, referred to as low-temperature methane conversion \cite{nozaki_non-thermal_2013}. Furthermore, methane plasmas are found in a variety of thin film applications, such as diamond synthesis by plasma-assisted vapour deposition \cite{kamo_diamond_1983}. Other applications range from modelling lightning in methane-containing atmospheres (such as Titan \cite{yair_study_2009, Kohn2019}) to studies of carbon-impurities inside a tokamak \cite{horton_atomic_1996}.

\subsection{Demands on cross sections}\label{sec:demands}

Theoretical and computational studies that underlie and enable the aforementioned applications all require a set of cross sections of electron collisions as model input. Although the requirements that are placed on a cross section data set can vary between applications, in general a set is required to be complete and consistent. Within the framework of low-temperature plasma modelling these properties are often defined 
according to Pitchford {\it et al.}~\cite{pitchford_comparisons_2013} as:
\begin{itemize}
    \item \textit{Complete} cross section sets accurately represent all electron energy and momentum losses as well as the electron number changing processes such as ionization and attachment.
    \item \textit{Consistent} cross section sets are able to reproduce measured values of swarm parameters within an order of ten percent, when used as an input to evaluate the electron energy distribution function from a Boltzmann solver.
\end{itemize}
Note that these definitions only apply to the behaviour of electron swarms. Other important demands on cross sections, such as the correct approximation of the production rates for chemical species, are not addressed.

When compiling a data set it is often found that experimental data alone is insufficient to ensure completeness and consistency, as data for crucial processes might be missing or measurements from different studies might disagree. The existence of such gaps in the literature can typically be attributed to the challenging nature of measurements for scattering processes such as: rotational and electronic excitation, dissociative attachment and, most notably, neutral dissociation \cite{buckman_atomic_2013}. Although theoretical cross section calculations can be used to supplement the experiments, such results are often constrained to specific energy ranges and are limited to simple molecules with low atom numbers. Within the framework of low-temperature plasma modelling a common method to overcome the limitations imposed by missing data is to fit presumably incorrect cross sections in order to have better agreement with measured swarm parameters \cite{petrovic2007kinetic}, c.f.\ the IST-Lisbon data set \cite{alves_ist-lisbon_2014}. Such data-fitting techniques are immensely enabling for their ability to produce consistent data sets in the absence of reliable measurements. However by fitting cross section data the scope of applicability of a data set is limited to describing the electron swarm behaviour of a plasma, as the rates of individual processes may have been altered significantly and the resulting cross section set can be non-unique \cite{petrovic2007kinetic, Crompton1994, song_recommended_2020}. In other words, plasma models using such \textit{swarm-fitted} data sets are not guaranteed to predict production rates of individual chemical species with a high degree of accuracy.

With an eye on accurately predicting the production of reactive species, it would be a highly attractive property for a cross section set to reproduce swarm parameters without the need for any fitting procedures. For such a set consistency is an inherent trait, i.e.\ independent of the limitations imposed by the swarm-fitting procedure. This would be especially attractive for applications that focus on the plasma-chemical activation of the gas, since the absence of a fitting procedure gives greater confidence that the individual cross sections are close to their `true' value. Moreover, an unfitted and consistent cross section set could be used in any plasma-modelling approach (e.g.\ hydrodynamic, multi-term Boltzmann or Monte-Carlo/PIC).

\subsection{Goal of the paper}
The goal of this paper is to derive cross sections for the neutral dissociation of the ground state of \CH{4}{} by electron impact. Secondly, we want to show that these cross sections in combination with data on other relevant scattering processes in \CH{4}{} produces a complete and consistent data set without the need for any data fitting. Our efforts are documented in two parts: in section \ref{sec:method}, we will review experimental and theoretical literature on the electron collision cross sections of \CH{4}{}. We highlight a gap in the literature corresponding to the neutral dissociation processes. In order to fill in this gap we propose a blend of empirical and analytical cross sections for the neutral dissociation processes in the energy range up to $100$~eV. In section \ref{sec:compare} we show that the addition of our cross sections to the recommendations of Song~\textit{et al.} \cite{song_cross_2015} produces a complete and consistent data set \textit{without} the need for any data-fitting techniques. By performing a Boltzmann analysis in pure methane we show that the agreement between calculated and measured swarm parameters is within error margins. 

Finally, in section \ref{sec:radicals} we compare the production of hydrogen species as given by our cross section set and the IST-Lisbon data set \cite{alves_ist-lisbon_2014}. The observed differences underline the issue regarding the non-uniqueness of swarm-fitted cross section sets.

\subsection{Relation to earlier work}
An extensive data evaluation regarding electron scattering with \CH{4}{} was published by Song~\textit{et al.} \cite{song_cross_2015}. In their work they recommend cross section values for most of the electron-neutral collisions: momentum-transfer \cite{allan_improved_2007, fedus_ramsauer-townsend_2014, sakae_scattering_1989, itikawa_63_2003, Kurachi_1990}, vibrational excitation \cite{Kurachi_1990}, ionization \cite{itikawa_51_2003} and dissociative electron attachment \cite{rawat_absolute_2008}. However, recommendations for the neutral dissociation processes have explicitly not been made due to inconsistencies in the available data. In section \ref{sec:compare} we demonstrate, by performing a Boltzmann analysis in pure methane, that simply neglecting these processes results in an ionization rate that is a factor ten larger than experimentally observed (This behaviour has also been documented in \cite{gadoum_set_2019}). The reason for this is that neutral dissociation processes are an important sink of electron energy that must be incorporated. 

Approximations for the missing cross sections of the electron impact dissociation processes are also presented by Gadoum and Benyoucef~\cite{gadoum_set_2019}. In essence, they employ a variation of the approximation technique formulated by Erwin and Kunc~\cite{erwin_electron-impact_2005, erwin_dissociation_2008}. The latter is also thoroughly discussed and evaluated in this study. The variant that Gadoum and Benyoucef~\cite{gadoum_set_2019} have used contains more fitting parameters in their low-energy approximation. Also by reordering the formulae, their variant requires the total (neutral and ionizing) cross sections into \CH{3}{} as an input parameter (which they have obtained from Motlagh and Moore \cite{motlagh_cross_1998}) instead of the total neutral dissociation. To avoid having to discuss two variants of the same approximation technique we have chosen to only include the original approximation technique formulated by Erwin and Kunc~\cite{erwin_electron-impact_2005, erwin_dissociation_2008} in our analysis.

Data for a wider range of hydrocarbon collision processes in a near-wall fusion plasma have been assembled by Janev and Reiter \cite{janev_collision_2002, reiter_hydrocarbon_2010}. The interest in that work is the complete breakdown chain of methane, ethane and propane, so including neutral and charged dissociation cross sections for electron impact on ${\rm C}_x{\rm H}_y$ with \mbox{$1\leq x\leq 3$} and \mbox{$1\le y\le 2x+2$}. Because of the paucity of data the emphasis in the work of Janev and Reiter, especially in the more recent work \cite{reiter_hydrocarbon_2010} for the case of neutral dissociation, is on the development of physically plausible functional forms for the cross sections for all target hydrocarbons. The data in \cite{janev_collision_2002, reiter_hydrocarbon_2010} are valuable and widely used for simulations of fusion plasma with carbon-based wall material where collisions involving many distinct hydrocarbon radical fragments are important. For our application to collisions with \CH{4}{} alone the data in \cite{janev_collision_2002, reiter_hydrocarbon_2010} are lacking validation and uncertainty estimates, so for us the starting point is Song \textit{et al.}~\cite{song_cross_2015} which we supplement with neutral dissociation cross sections validated to swarm data.

\section{Compilation of unfitted cross sections for neutral dissociation of \CH{4}{}}\label{sec:method}
We will start by evaluating the literature regarding neutral dissociation. We address the same inconsistencies as were found by Song~\textit{et al.}  \cite{song_cross_2015}, but for energies as high as $100$~eV. We then proceed by formulating our approximation for the cross sections of this process. We will lay an emphasis on the energy range of up to $100$~eV, relevant for low temperature plasmas. Note that some of our proposed cross sections extend up to $500$~eV, however such high energies are not shown because they have a negligible contribution on the computation of swarm parameters in numerical swarm experiments, which we will use to evaluate these approximations in section \ref{sec:compare}.

\subsection{Neutral dissociation of \CH{4}{}}\label{sec:literaturereview}

The dissociation processes of methane generally occur through electronic excitation of the molecule to an intermediate state\cite{ziolkowski_modeling_2012}. All of the electronically excited states of methane are short-lived and are dissociative or subject to auto-ionization, hence the intermediate electronic excited state can generally be omitted from consideration \cite{song_cross_2015}. For the excitations that lead to neutral dissociation several channels have to be considered:
\begin{numcases}{\text{e + \CH{4}{}} \rightarrow \text{e + \CH{4}{*}}\rightarrow}
    \text{e + \CH{3}{} + \dots } \label{eq:channelCH3}\\ 
    \text{e + \CH{2}{} + \dots} \label{eq:channelCH2}\\
    \text{e + \CH{}{} + \dots}\label{eq:channelCH} \\
    \text{e + C + \dots}\label{eq:channelC}
\end{numcases}
The cross sections of these neutral dissociation processes are denoted by $\sigma_{i}$ with $i$ representing the particular dissociated methane fragment: \CH{3}{}, \CH{2}{}, etc.
\begin{figure*}
    \centering
    \includegraphics[width=\textwidth]{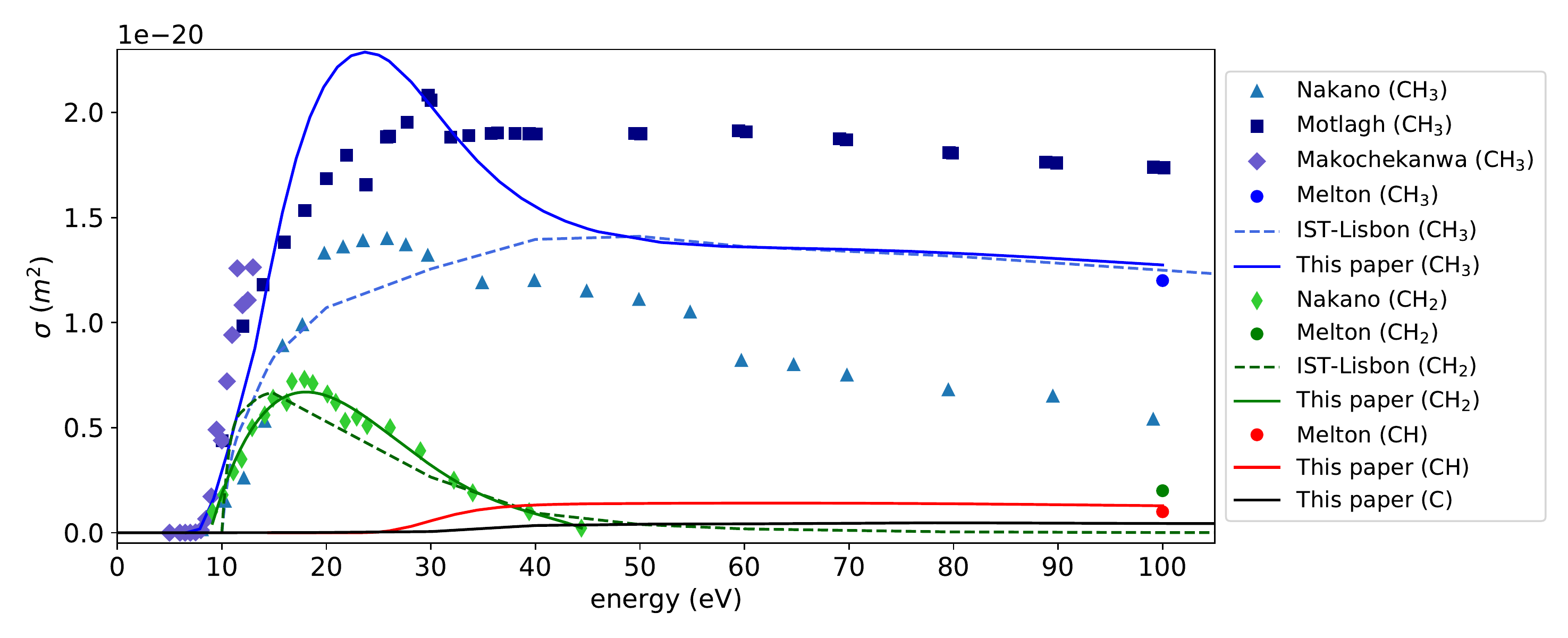}
    \caption{An overview of experimental values for neutral dissociation cross sections of each channel together with the fitted values from IST-Lisbon \cite{alves_ist-lisbon_2014} and our proposed values, within the considered energy range up to $100$~eV. The shown measurements are from: Nakano~\textit{et al.} \cite{nakano_electron-impact_1991-1, nakano_electron-impact_1991}, Motlagh and Moore~\cite{motlagh_cross_1998}, Makochekanwa~\textit{et al.}~\cite{makochekanwa_experimental_2006} and Melton and Rudolph~\cite{melton_radiolysis_1967}. Note that these measurements do not agree with each other.}
    \label{fig:CS_overview}
\end{figure*}

The body of literature regarding the neutral dissociation cross sections is sparse. For instance, no direct measurements below $100$~eV exist for $\sigma_{\CH{}{}}$ and $\sigma_\text{C}$. However, cross sections for the neutral dissociation into specific excited states, e.g.\ CH($A^2\Delta$) and CH($B^2\Sigma^-$), have been determined by {\v{S}}a{\v{s}}i{\'{c}}~\textit{et al.}~\cite{Sasic2004}.  For the remaining dissociation processes, the experimental observations are in disagreement and theoretical results are only available for a narrow energy range of $10$~eV to $16.5$~eV. In figure \ref{fig:CS_overview} we have shown a selection of the experimental results evaluated by Song~\textit{et al.} of $\sigma_{\CH{3}{}}$ and $\sigma_{\CH{2}{}}$ for electron energies up to $100$~eV. The relative experimental uncertainty accompanying these measurements are $\pm100\%$ for Nakano~\textit{et al.}  \cite{nakano_electron-impact_1991-1, nakano_electron-impact_1991}, $\pm30\%$ for Motlagh and Moore\ \cite{motlagh_cross_1998} and $\pm20\%$ for Makochekanwa~\textit{et al.} \cite{makochekanwa_experimental_2006}. In this figure we have also supplied the fitted cross sections from the IST-Lisbon data set \cite{alves_ist-lisbon_2014}, our recommendations which are derived at the end of this section and the isolated measurements of Melton and Rudolph \cite{melton_radiolysis_1967} for $\sigma_{\CH{3}{}}$, $\sigma_{\CH{2}{}}$ and $\sigma_{\CH{}{}}$ at $100$~eV. No measure of uncertainty was supplied for the measurements by Melton and Rudolph.

\begin{figure*}
\centering
\begin{minipage}[t]{.49\textwidth}
    \includegraphics[width=\textwidth]{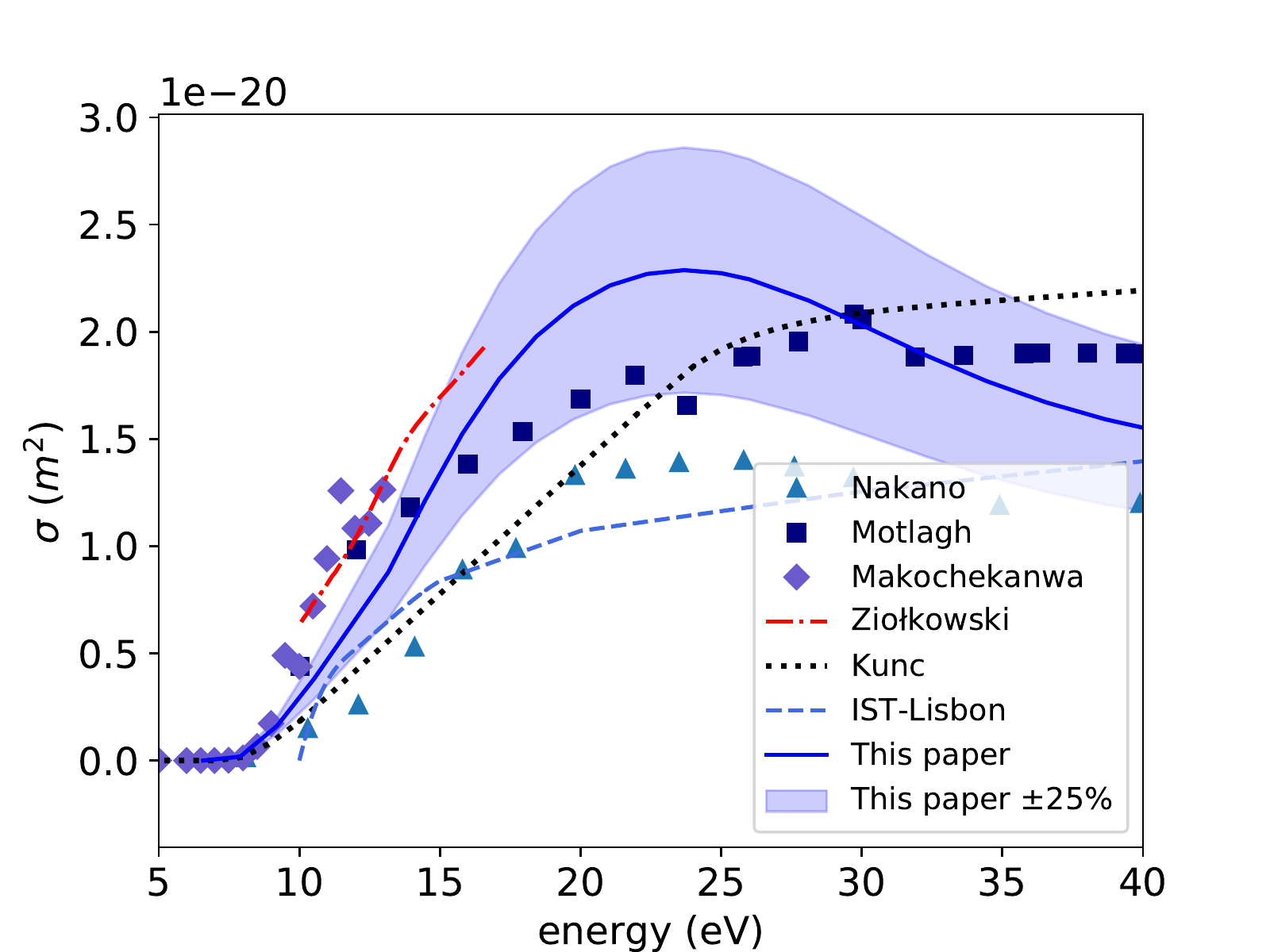}
    \caption{A zoom on the low-energy range of the cross sections of neutral dissociation into \CH{3}{} together with our recommendations (including a $\pm 25\%$ deviation). The shown values are obtained \textit{experimentally}: Nakano~\textit{et al.} ~\cite{nakano_electron-impact_1991-1, nakano_electron-impact_1991}, Motlagh and Moore~\cite{motlagh_cross_1998} and Makochekanwa~\textit{et al.} ~\cite{makochekanwa_experimental_2006}, \textit{theoretically}: Zio\l kowski~\textit{et al.} ~\cite{ziolkowski_modeling_2012}, by \textit{semi-empirical approximations}: Erwin and Kunc~\cite{erwin_electron-impact_2005, erwin_dissociation_2008} or by \textit{swarm fitting}: IST-Lisbon \cite{alves_ist-lisbon_2014}. Note the agreement between Zio\l kowksi and Makochekanwa.}
    \label{fig:CS_CH3}
\end{minipage}\qquad
\begin{minipage}[t]{.49\textwidth}
    \includegraphics[width=\textwidth]{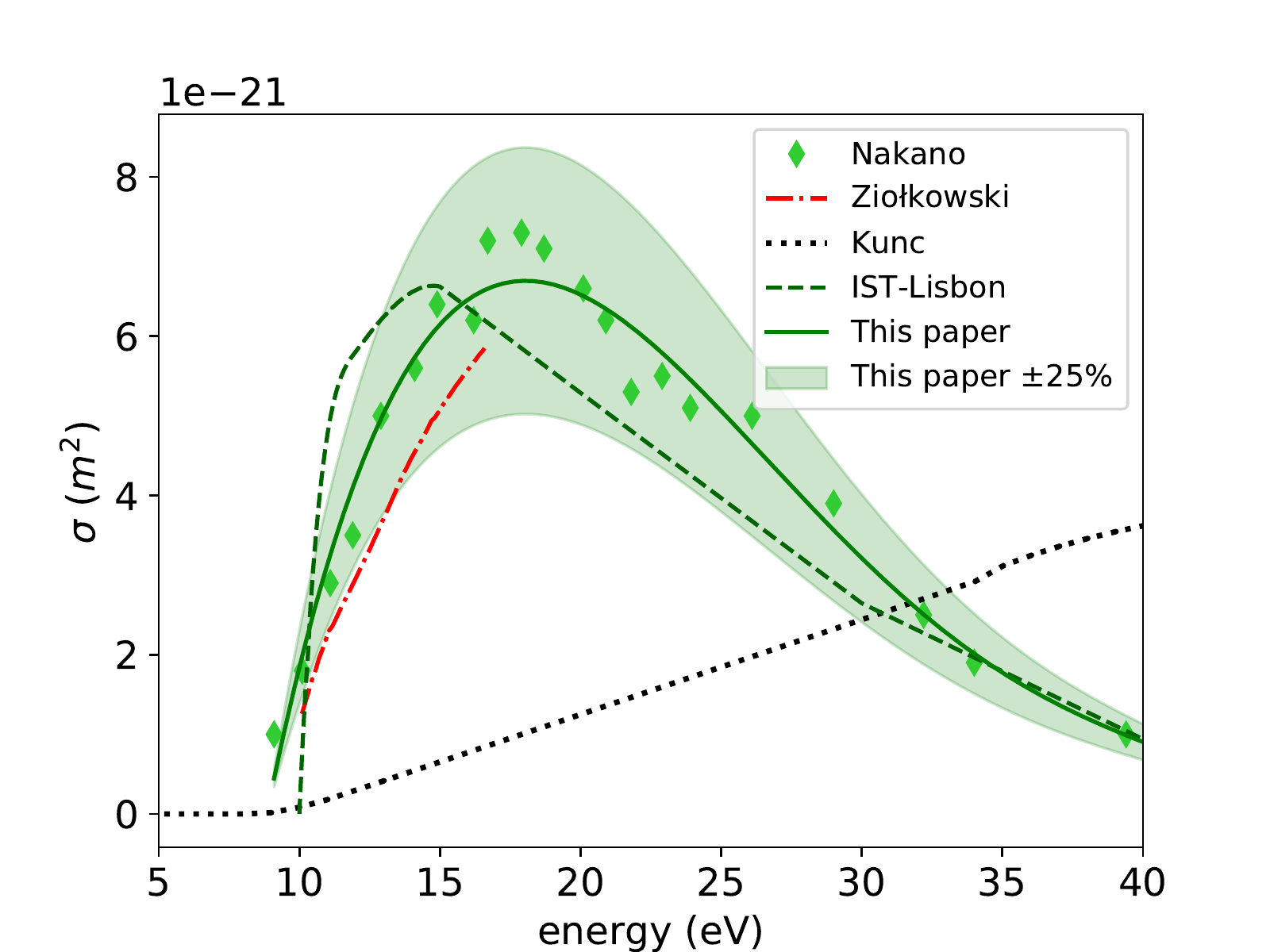}
    \caption{A zoom on the low-energy range of the cross sections of neutral dissociation into \CH{2}{} together with our recommendations (including a $\pm 25\%$ deviation). The shown values are obtained \textit{experimentally}: Nakano~\textit{et al.} ~\cite{nakano_electron-impact_1991-1, nakano_electron-impact_1991} \textit{theoretically}: Zio\l kowski~\textit{et al.} ~\cite{ziolkowski_modeling_2012}, by \textit{semi-empirical approximations}: Erwin and Kunc~\cite{erwin_electron-impact_2005, erwin_dissociation_2008} or by \textit{swarm fitting}: IST-Lisbon \cite{alves_ist-lisbon_2014}. Note the agreement between Zio\l kowksi and Nakano.}
    \label{fig:CS_CH2}
\end{minipage}
\end{figure*}

In figure \ref{fig:CS_CH3} we zoom in and compare the values of $\sigma_{\CH{3}{}}$ up to $40$~eV from the experimental observations mentioned above with the results from theoretical calculations by Zio\l kowski~\textit{et al.} \cite{ziolkowski_modeling_2012} and with the approximations from Erwin and Kunc\ \cite{erwin_electron-impact_2005, erwin_dissociation_2008}. Note the agreement between recent experimental and theoretical results from Makochekanwa~\textit{et al.} \cite{makochekanwa_experimental_2006} and Zio\l kowski~\textit{et al.} \cite{ziolkowski_modeling_2012} which shows a sharp increasing cross section in the near-threshold region. Based on this agreement and the fundamental nature of their work, Zio\l kowski~\textit{et al.}\cite{ziolkowski_modeling_2012} conclude that their prediction and the measurements of Makochekanwa~\textit{et al.} \cite{makochekanwa_experimental_2006} of $\sigma_{\CH{3}{}}$ are more reliable than the results of Nakano~\textit{et al.} \cite{nakano_electron-impact_1991-1, nakano_electron-impact_1991} and Erwin and Kunc \cite{erwin_electron-impact_2005, erwin_dissociation_2008}. Furthermore, Zio\l kowski~\textit{et al.} \cite{ziolkowski_modeling_2012} observe that within their considered energy range, $10$ to $16.5$~eV, their predictions match with Motlagh and Moore \cite{motlagh_cross_1998}.
However when transposing their measured relative cross sections to an absolute scale, Motlagh and Moore only considered neutral dissociation into \CH{3}{}. This means that the contributions due to $\sigma_{\CH{2}{}}$, $\sigma_{\CH{}{}}$ and $\sigma_{\text{C}}$ are neglected. Although these cross sections are not known exactly we estimate, based on the measurements of Nakano~\textit{et al.}  \cite{nakano_electron-impact_1991-1, nakano_electron-impact_1991}, that the cross sections for $\sigma_{\CH{2}{}}$ are considerable in the region between the threshold energy (which can be estimated to lie around $7.5$~eV) and $20$~eV. For this reason we do not use the measured cross sections for neutral dissociation into \CH{3}{} from Motlagh and Moore~\cite{motlagh_cross_1998}.

A zoomed-in view of $\sigma_{\CH{2}{}}$ is shown in figure \ref{fig:CS_CH2}. In this case the only experimental observation for energies below $40$~eV are reported by Nakano~\textit{et al.}  \cite{nakano_electron-impact_1991-1, nakano_electron-impact_1991}. Contrary to their results for $\sigma_{\CH{3}{}}$, the values of $\sigma_{\CH{2}{}}$ are in excellent agreement with the theoretical predictions from Zio\l kowski~\textit{et al.} \cite{ziolkowski_modeling_2012} in the near-threshold energy region. Moreover, the approximation by Erwin and Kunc \cite{erwin_dissociation_2008} deviates significantly from the aforementioned results, as it does not portray the sharp rise for low energies. This qualitative difference might be attributed to the absence of any data calibration, aside from fixing a threshold energy, of the low-energy ($<50$~eV) approximation of Erwin and Kunc \cite{erwin_dissociation_2008}. For these reasons, Zio\l kowski~\textit{et al.} \cite{ziolkowski_modeling_2012} conclude that their results and the measurements of Nakano~\textit{et al.} \cite{nakano_electron-impact_1991-1, nakano_electron-impact_1991} for the dissociation into \CH{2}{} are more reliable.

\subsection{Our proposed cross sections}
In the previous section it was discussed, relying on the advancements regarding neutral dissociation cross sections in the low-energy regime \cite{makochekanwa_experimental_2006, ziolkowski_modeling_2012}, that the only available measurements for dissociation into \CH{3}{} across a wide energy range (i.e.\ Motlagh and Moore \cite{motlagh_cross_1998} and Nakano~\textit{et al.}  \cite{nakano_electron-impact_1991-1, nakano_electron-impact_1991}) are unsatisfactory.

For this reason we resort to an alternative method to obtain cross sections for neutral dissociation into \CH{3}{}, following Janev and Reiter \cite{janev_collision_2002} and Erwin and Kunc \cite{erwin_electron-impact_2005} with support from measurements of Winters \cite{winters_dissociation_1975}. Winters \cite{winters_dissociation_1975} observed that for energies above $50$\,eV the total dissociation cross section is split equally between neutral and ionizing dissociation, suggesting a common mechanism. Janev and Reiter \cite{janev_collision_2002} describe the common mechanism as: “[...] excitation of a dissociative state which lies in the ionization continuum. Autoionization of this state leads to dissociative ionization, while its survival leads to dissociation to neutrals.” They conclude from this that cross section branching ratios within the neutral dissociation channel should match cross section branching ratios within the ionized dissociation channel. Erwin and Kunc \cite{erwin_electron-impact_2005} treat these branching ratios in a similar manner. 
Therefore, consistent with Janev and Reiter \cite{janev_collision_2002} and with Erwin and Kunc \cite{erwin_electron-impact_2005} we chose to employ for electron impact energies above the threshold for ionizing dissociation the following functional approximation for $\sigma_{\CH{3}{}}$:
\begin{align}\label{eq:sigCH3Winters}
    \sigma_{\CH{3}{}} &= \frac{\sigma_\text{ND}}{\sigma_\text{ID}} (\sigma_{[\CH{3}{+}+\text{H}]}+\sigma_{[\CH{3}{}+\text{H}^+]}),\nonumber\\
     & \text{for } \epsilon \geq \epsilon_c,
\end{align}
with $\sigma_\text{ID}$ the cross section for total ionizing dissociation, $\sigma_\text{ND}$ the cross section for total neutral dissociation, $\sigma_{[\CH{3}{+}+\text{H}]}$ and $\sigma_{[\CH{3}{}+\text{H}^+]}$ correspond to cross sections of specific ionizing dissociation by electron impact and $\epsilon_c$ is the lowest energy at which ``reliable" experimental cross sections are available. For ionization processes we adopted the cross sections reported by Lindsay and Mangan~\cite{itikawa_51_2003}, as is recommended in Song \textit{et al.}~\cite{song_cross_2015}. Furthermore, the value of $\sigma_\text{ND}$ can be obtained by subtracting the total ionizing dissociation from the total dissociation:
\begin{equation}
    \sigma_\text{ND} = \sigma_\text{TD} - \sigma_\text{ID},
\end{equation}
with $\sigma_\text{TD}$ the cross section for the total dissociation reported by Winters~\cite{winters_dissociation_1975}. Fitting functions reported in Shirai \textit{et al.}~\cite{shirai_analytic_2002} were used for $\sigma_\text{TD}$, all ionizing dissociation cross sections, and the dissociative electron attachment cross sections. More details on these fitting functions and their parameters can be found in \ref{app:fit}. We can recover the initial observation of Winters by taking $\sigma_\text{ND} = \sigma_\text{ID}$, which generally holds for energies above $50$~eV. Note that the approximation from equation \ref{eq:sigCH3Winters} only holds for energies above the threshold energies of the corresponding ionizing dissociation reactions.

\begin{table*}
    \centering
    \begin{tabular}{r||*{6}{c|}}
         & $\epsilon_\text{ND}$ (eV) & $\epsilon_b$ (eV) & $\epsilon_c$ (eV) & $\sigma^{(2)}_i(\epsilon_c)$ (m$^2$) & $a$ (m$^2$) & ~$p$~\\
         \hline
        $\sigma_{\CH{3}{}}$ & $7.5$ & $10.5$ & $13.16$ & $8.8 \cdot 10^{-21}$ & $1.5 \cdot 10^{-20}$ & $1.5$ \\
        $\sigma_{\CH{}{}}$ & $15.5$ & $18.5$ & $22.37$ & $2.9 \cdot 10^{-27}$ & $1.3 \cdot 10^{-26}$ & $1.6$ \\
        $\sigma_\text{C}$ & $15.5$ & $18.5$ & $22.37$ & $6.8 \cdot 10^{-29}$ & $3.1 \cdot 10^{-28}$ & $1.6$ \\
    \end{tabular}
    \caption{The parameters used for the low-energy approximations of our proposed cross sections.}
    \label{tab:LowEnergyParameters}
\end{table*}

However, neutral dissociation reactions have a lower threshold energy than their respective ionizing reactions and therefore also occur at energies below the ionization threshold. Thus for energies below the respective ionizing dissociation thresholds we apply the low-energy approximation method of Erwin and Kunc. Here we only present the final result applied to $\sigma_\text{\CH{3}{}}$ used in our work, for a detailed discussion we refer to the original work \cite{erwin_dissociation_2008}. In this method the below-ionization energy range is divided in a near-threshold range, ${\epsilon_\text{ND}\leq\epsilon\leq\epsilon_b}$, and a linear-growth range ${\epsilon_b\leq\epsilon\leq\epsilon_c}$. Here $\epsilon_\text{ND}$ represents the threshold energy for neutral dissociation, $\epsilon_b$ represents the energy value separating the near-threshold range from the linear growth range. Then the near-threshold cross section is given by:
\begin{align}\label{eq:sigCH3Kunc1}
    \sigma_{\CH{3}{}} &= a\left(\frac{\epsilon}{\epsilon_\text{ND}}-1\right)^p,\nonumber\\
    &\text{ for } \epsilon_\text{ND}\leq\epsilon\leq\epsilon_b,
\end{align}
with $a$ and $p$ positive constants. For the linear-growth range the cross section are blended with the relation from equation \eqref{eq:sigCH3Kunc1} as follows:
\begin{eqnarray}\label{eq:sigCH3Kunc2}
    \sigma_{\CH{3}{}}&=& \sigma_{\CH{3}{}}^{(1)}(\epsilon_b)  + \frac{ \sigma_{\CH{3}{}}^{(2)}(\epsilon_c) -  \sigma_{\CH{3}{}}^{(1)}(\epsilon_b)}{\epsilon_c - \epsilon_b}(\epsilon-\epsilon_b),    
    \nonumber\\
    &&\mbox{for }\epsilon_b\leq\epsilon\leq\epsilon_c,
\end{eqnarray}
with blending-parameter $\sigma_{\CH{3}{}}^{(1)}(\epsilon_b)$ representing the value of the cross section evaluated at $\epsilon_b$ as calculated from equation \eqref{eq:sigCH3Kunc1}, and $\sigma_{\CH{3}{}}^{(2)}(\epsilon_c)$ the value of the experimentally-obtained cross section at corresponding energy $\epsilon_c$ corresponding to equation \eqref{eq:sigCH3Winters}. As can be seen, equations \eqref{eq:sigCH3Winters}-\eqref{eq:sigCH3Kunc2} determine the cross section for neutral dissociation into \CH{3}{} for the whole energy range.

The cross sections for the neutral dissociation into \CH{}{} and C are obtained analogously. The values for the parameters $\epsilon_\text{ND}$, $\epsilon_b$, $\epsilon_c$, $\sigma_i^{(2)}(\epsilon_c)$, $a$ and $p$ used in our work are given in table \ref{tab:LowEnergyParameters}. The parameters for $\sigma_\text{C}$ are not covered by Erwin and Kunc \cite{erwin_dissociation_2008}, but here they are obtained following the same reasoning for the parameters of \CH{}{}. 

Although the approximations defined in equations \eqref{eq:sigCH3Winters}-\eqref{eq:sigCH3Kunc2} can be used for any of the neutral dissociation channels, they are only used for $\sigma_{\CH{3}{}}$, $\sigma_{\CH{}{}}$ and $\sigma_\text{C}$. For the remaining neutral dissociation process, $\sigma_{\CH{2}{}}$, relying on the experimental observations by Nakano~\textit{et al.}  \cite{nakano_electron-impact_1991-1, nakano_electron-impact_1991} is preferred over the application of a similar approximation, due to the agreement with theoretical predictions by Zio\l kowski~\textit{et al.}  \cite{ziolkowski_modeling_2012}. Thus, in this study we take $\sigma_{\CH{2}{}}$ to be given by a fourth-order polynomial fit through the measurements of Nakano~\textit{et al.}. We refer to \ref{app:fit} for the fitting parameters.

\begin{figure}
    \centering
    \includegraphics[width=0.49\textwidth]{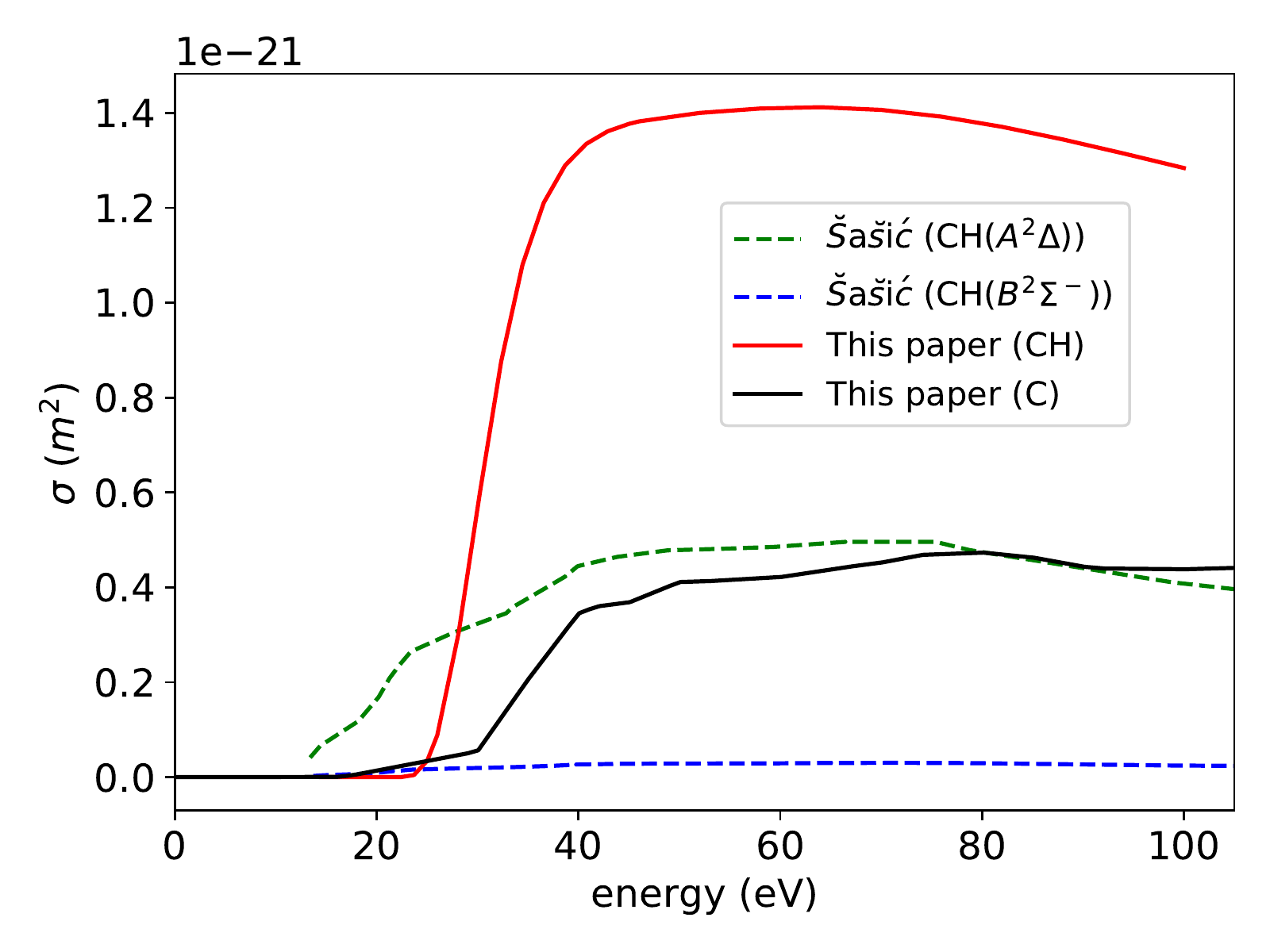}
    \caption{The cross sections $\sigma_\text{\CH{}{}}$ and $\sigma_\text{C}$ used in our work alongside the experimentally-derived cross sections into specific excited fragments by {\v{S}}a{\v{s}}i{\'{c}}~\textit{et al.}~\cite{Sasic2004}. The latter should always be smaller than $\sigma_\text{\CH{}{}}$. This holds in general, aside from a small discrepancy in the vicinity of the threshold, i.e.\ below $25$\,eV.}
    \label{fig:CS_CH_C}
\end{figure}

Our proposed cross sections for the neutral dissociation processes are shown in figures \ref{fig:CS_overview}-\ref{fig:CS_CH2}. For $\sigma_\text{\CH{3}{}}$ the qualitative trend of our proposed cross section is similar to the results from Makochekanwa~\textit{et al.} \cite{makochekanwa_experimental_2006} and Zio\l kowski~\textit{et al.} \cite{ziolkowski_modeling_2012} in the near-threshold energy region, although it appears shifted to higher energies by around $1.5$~eV. Our proposed cross sections have a maximum value of ${2.29\cdot10^{-20}~\text{m}^2}$ at $24$~eV, which is higher than any of the experimental results. After attaining this maximum the value decays and eventually agrees with the isolated measurement of Melton and Rudolph \cite{melton_radiolysis_1967} at $100$~eV. Note also that for energies above $50$\,eV our proposed value corresponds to the fitted values from the IST-Lisbon set \cite{alves_ist-lisbon_2014}. Our proposed values for $\sigma_\text{\CH{2}{}}$, based on the measurements of Nakano~\textit{et al.} \cite{nakano_electron-impact_1991-1, nakano_electron-impact_1991}, vanish for energies above $45$~eV. This contradicts with the measurements from Melton and Rudolph \cite{melton_radiolysis_1967}, which suggest that the cross section should be around ${1.95\cdot10^{-21}~\text{m}^2}$ at $100$~eV. This difference is also recognized by Nakano~\textit{et al.}~\cite{nakano_electron-impact_1991-1, nakano_electron-impact_1991}. To the best of our knowledge, there is currently no straightforward method to reconcile these two observations. Moreover, our proposed cross sections agree (qualitatively) with the fitted counterparts from the IST-Lisbon set \cite{alves_ist-lisbon_2014}, although the latter appears to have shifted the peak to lower energies by approximately $4$\,eV.  For $\sigma_\text{\CH{}{}}$ and $\sigma_\text{C}$ there is little literature to compare with aside from observing that our approximation of $\sigma_\text{\CH{}{}}$ agrees with the isolated measurement of Melton and Rudolph at $100$~eV. Furthermore, we can compare our values of $\sigma_\text{\CH{}{}}$ with the results for the neutral dissociation into the excited fragments CH($A^2\Delta$) and CH($B^2\Sigma^-$) which have been determined by {\v{S}}a{\v{s}}i{\'{c}}~\textit{et al.}~\cite{Sasic2004}. It should hold that the dissociation into specific excited fragments is lower than $\sigma_\text{\CH{}{}}$. As shown in figure \ref{fig:CS_CH_C} this behaviour generally holds. Only in the vicinity of the threshold, i.e.\ below $25$\,eV, do we observe that the cross sections for dissociation into excited fragments are higher than our proposed value. However, this discrepancy is small compared to the dominant inelastic scattering processes and will therefore be negligible within the context of the swarm experiments that are presented in the following sections.

\section{Comparison of calculated and measured swarm parameters} \label{sec:compare}
Within the framework of low-temperature plasma modelling, a computation of swarm parameters is performed routinely, typically for reduced electric fields ($E/N$) between $0.1$~Td and $1000$~Td, with $E$ representing the electric field and $N$ the number density of the gas. In a fluid description of electron swarms (e.g.\ \cite{salabas_two-dimensional_2002} and references therein) the electrons are described by their density only and this density obeys a reaction-drift-diffusion equation governed by swarm parameters: diffusion coefficient $D$, mobility $\mu$ and by coefficients for ionization $\alpha$ and attachment $\eta$. Moreover, the characteristic energy $D/\mu$, reduced mobility $\mu N$ and reduced Townsend ionization coefficient $\alpha/N$ are functions of the reduced electric field $E/N$ only (for a not too large electric fields). These swarm parameters can be obtained, given the gas composition and a cross section data set, by solving the Boltzmann equation \cite{frost_rotational_1962}.

In this section we will use the difference between computed and measured swarm parameters as an implicit metric to evaluate the cross sections for neutral dissociation in conjunction with the recommendations from Song~\textit{et al.}  \cite{song_cross_2015} (neglecting rotational excitations since these are already accounted for in the elastic momentum-transfer cross section). Note that explicit evaluation of the cross sections for neutral dissociation processes is not possible due to disagreement in the available literature, as was shown in section \ref{sec:literaturereview}. On the other hand, the swarm parameters of a methane plasma are well-known, with the exception of the attachment coefficient. This can be seen from a compilation made in Alves~\cite{alves_ist-lisbon_2014} of measurements containing observations for reduced mobility $\mu N$, characteristic energy $D/\mu$, and the reduced Townsend ionization coefficient $\alpha/N$ \cite{al-amin_electron_1985, cochran_diffusion_1962, cottrell_drift_1965, davies_measurements_1989, heylen_ionization_1975, hunter_electron_1986, lakshminarasimha_ratio_1977, lin_moment_1979, millican_electron_1987, pollock_momentum_1968, shimozuma_measurement_1981}. Assuming that the recommendations by Song~\textit{et al.}  have a sufficiently low error margin, any disagreement between calculated and measured swarm parameters must imply that the remaining cross sections, i.e.\ the neutral dissociation processes, are inaccurate. We will compare bulk swarm parameters as computed by 
a Monte-Carlo solver \cite{particle_swarm} based on the modelling framework presented in \cite{Teunissen2016}. The simulations are performed at standard temperature and pressure. We emphasize that we show the bulk coefficients and that the characteristic energy is based on the transversal diffusion coefficient.
The swarm parameters have been computed for four cross section data sets: 
\begin{enumerate}
    \item the \textit{swarm-fitted} IST-Lisbon database \cite{alves_ist-lisbon_2014},\label{it:Lisbon}
    \item the recommendations by Song~\textit{et al.}  \cite{song_cross_2015} (lacking any neutral dissociation process),\label{it:Song}
    \item the recommendations by Song~\textit{et al.}  in conjunction with the original approximations by Erwin and Kunc \cite{erwin_electron-impact_2005, erwin_dissociation_2008} for neutral dissociation,\label{it:SongKunc}
    \item the recommendations by Song~\textit{et al.}  in conjunction with our approximations for neutral dissociation.\label{it:Ours}
\end{enumerate}
Moreover, for data set \eqref{it:Ours} we have included the effect of varying our proposed cross sections for neutral dissociation by $\pm25\%$. This results in an upper and lower bound for the reproduced swarm parameters. These bounds define a range which we will refer to as the sensitivity interval. This interval is included to illustrate the effect that possible errors on the cross sections for neutral dissociation might impose on the computed swarm parameters.

\begin{figure*}
    \centering
    \begin{minipage}[t]{.9\textwidth}
    \includegraphics[width=\textwidth]{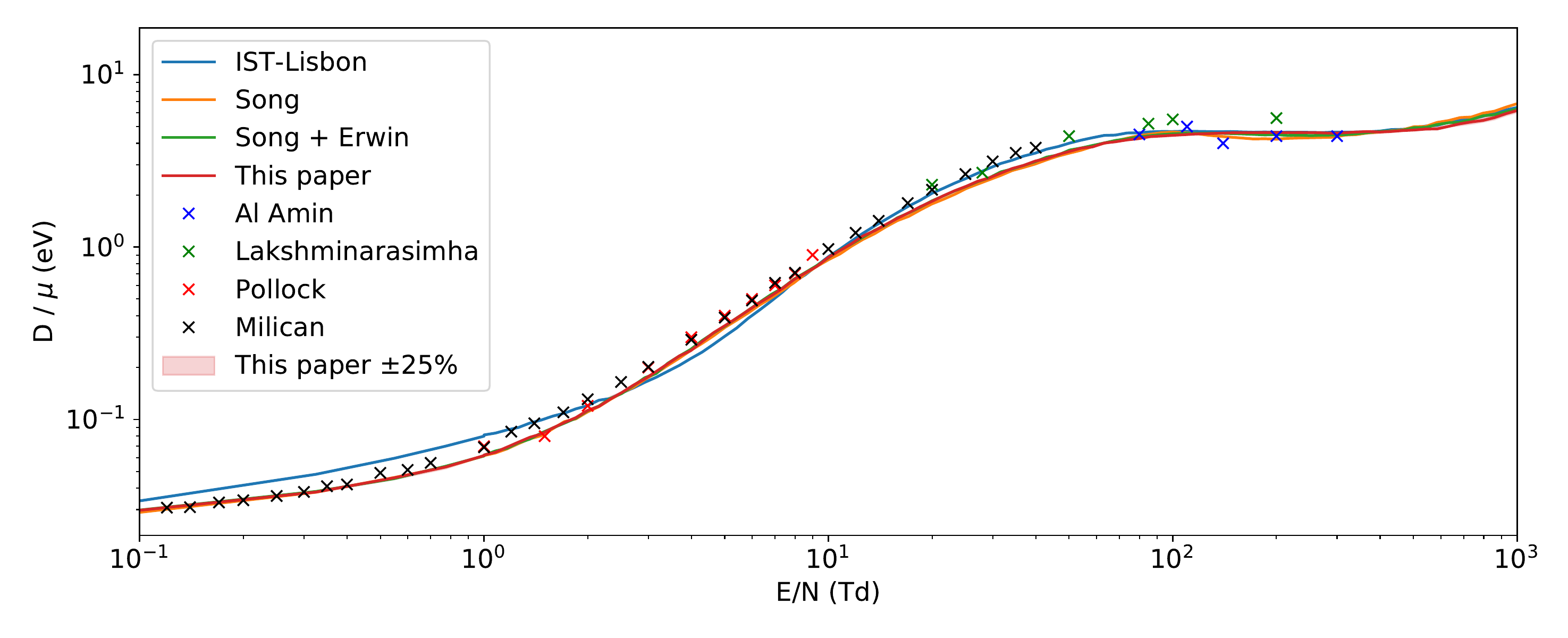}
    \subcaption{characteristic energy $D/\mu$}
    \label{fig:char_en}
    \end{minipage}
    \begin{minipage}[t]{.9\textwidth}
    \includegraphics[width=\textwidth]{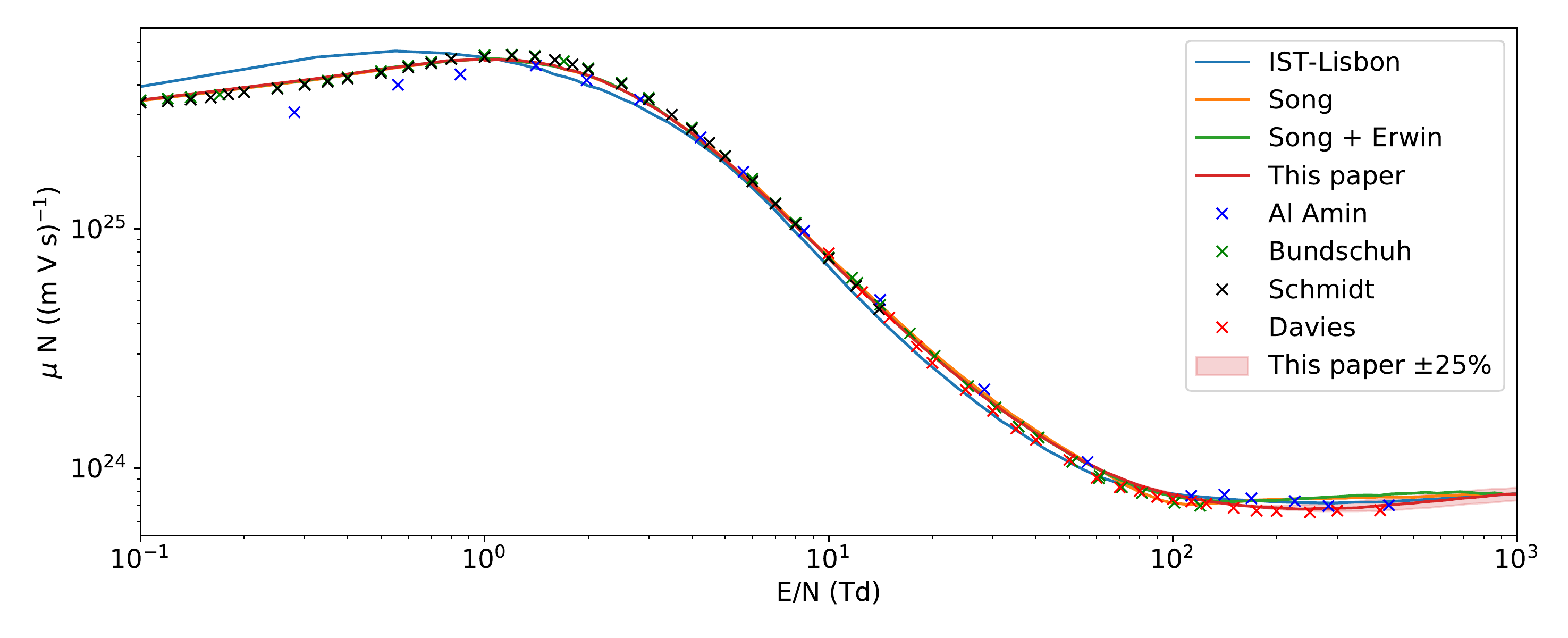}
    \subcaption{reduced mobility $\mu N$}
    \label{fig:mu}
    \end{minipage}
    \begin{minipage}[t]{.9\textwidth}
    \includegraphics[width=\textwidth]{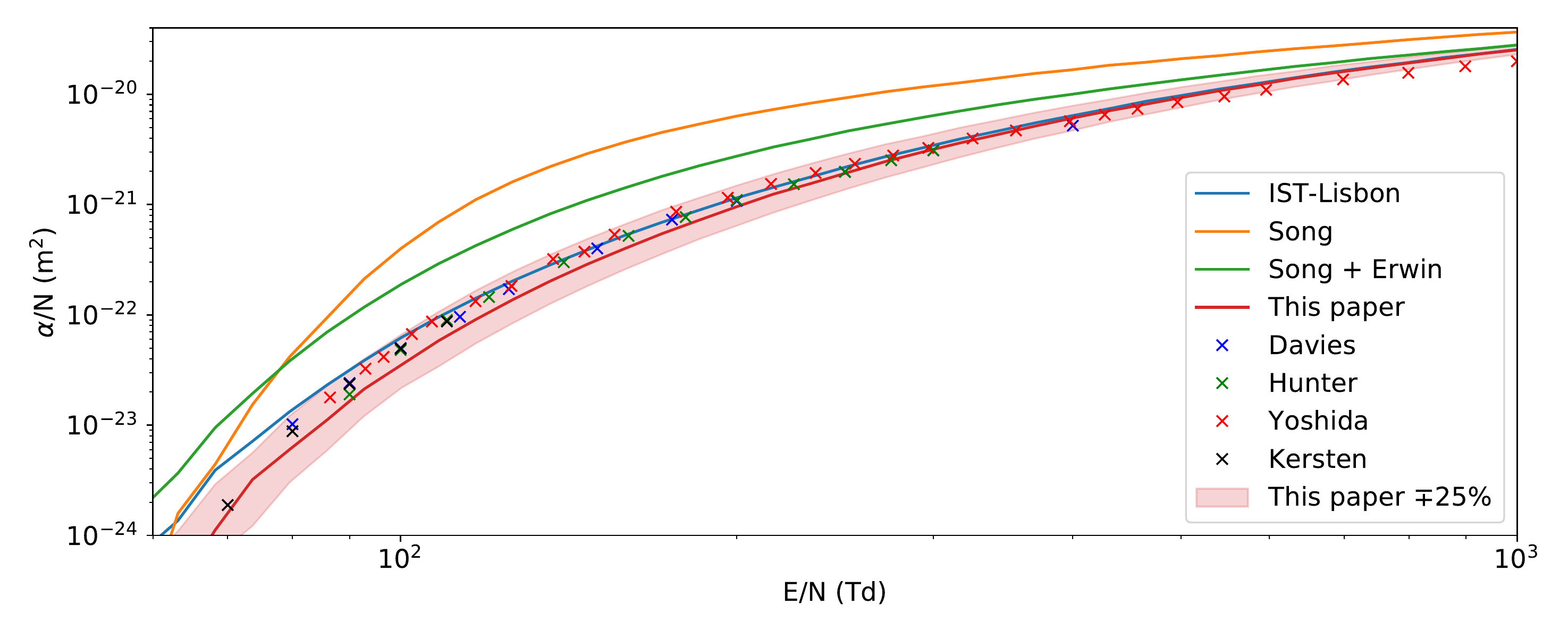}
    \subcaption{reduced Townsend ionization coefficient $\alpha/N$}
    \label{fig:alpha}
    \end{minipage}
    \caption{Measured and calculated values of the swarm parameters in pure methane. The shaded red region corresponds to the sensitivity interval, obtained by including a $\pm25\%$ deviation on the neutral dissociation processes. Overestimating the neutral dissociation leads to underestimating the ionization, and vice versa, hence the use of the `$\mp$' sign in \ref{fig:alpha}. Our cross section set reproduces swarm parameters within a few tens of percent.}
\end{figure*}

In figures \ref{fig:char_en}-\ref{fig:alpha} we have respectively shown the characteristic energy, mobility and ionization from numerical and experimental studies on a double logarithmic scale. All of the considered data sets reproduce the characteristic energy within an error margin of $20\%$ and the mobility within an error margin of $7.5\%$, as can be seen in figures \ref{fig:char_en} and \ref{fig:mu}. One exception to this can be found at reduced electric fields below $10$ Td. Where data set \eqref{it:Lisbon} exhibits deviations from the measured values (and the respective error margins) of the characteristic energy ($25\%$) and the mobility ($15\%$ between $1$\,Td and $10$\,Td and $30\%$ below $1$\,Td). On the other hand the ionization coefficient, in figure 6, varies strongly between different data sets. Data set \eqref{it:Song} overshoots the measured values by as much as a factor of ten. Clearly the neutral dissociation of methane plays a vital role in determining the electron number changing processes and must be incorporated. Even though adding the cross sections for neutral dissociation from Erwin and Kunc reduces this discrepancy, the corresponding data set \eqref{it:SongKunc} still exhibits notable discrepancies with measured ionization coefficients. Given the high values of the ionization coefficient it appears that data sets \eqref{it:Song} and \eqref{it:SongKunc} critically underestimate the sinks for electron energy. This can be explained by considering that the underestimation of energy losses means that an electron is more likely to obtain energies above the ionization potential and subsequently the rate of ionization increases. By inspecting figures \ref{fig:CS_CH3} and \ref{fig:CS_CH2}, one can observe that for electron energies below $25$~eV the values from Erwin and Kunc are lower than (most of) the other reported values.
This behaviour is especially pronounced for dissociation into \CH{2}{}. For swarm experiments in general, the electrons in this energy regime play a dominant role in determining swarm parameters as the abundance of electrons typically reduces strongly for higher electron energies. Therefore the effect of omitting or underestimating the neutral dissociation processes as is done in data sets \eqref{it:Song} and \eqref{it:SongKunc} can be expected to introduce large discrepancies in the computed ionization coefficient, as is also seen in figure \ref{fig:alpha}.

Such an overestimation of the ionization coefficient is not present for the other considered data sets. The swarm-fitted data set \eqref{it:Lisbon} reproduces the ionization coefficient with a maximum deviation of $25\%$ in the region between $100$\,Td and $800$\,Td. However below $100$\,Td the deviations starts to increase. For instance, at $90$\,Td this deviation exceeds $40\%$. The large accuracy between $100$\,Td and $800$\,Td is to be expected from data sets which employ fitting procedures to ensure completeness and consistency. At $1000$\,Td the deviation is around $35\%$. The reproduction of the ionization coefficient is also observed for our approximations in conjunction with Song~\textit{et al.}, data set \eqref{it:Ours}, with a maximum deviation up to $35\%$ (at $100$\,Td). This is somewhat larger than observed for data set \eqref{it:Lisbon}. For reduced electric fields below $100$\,Td our reduced Townsend ionization coefficient is closer to measurements than data set \eqref{it:Lisbon}. Notably, up to $500$\,Td it can be observed that our reduced Townsend ionization coefficient is consistently lower than experimentally observed ionization coefficient. This indicates, if one assumes that the ionizing cross sections are sufficiently accurate, that the sum of all non-ionizing inelastic cross sections used here is probably an overestimation.

Furthermore, from the sensitivity interval corresponding to data set \eqref{it:Ours} we can conclude that the reproduction of characteristic energy and mobility is almost completely independent of the neutral dissociation processes. In contrast, the sensitivity interval for the ionization coefficient shows a significant spread. This again underlies that neutral dissociation processes are an important electron energy sink, at least within the context of low-temperature plasmas. Moreover, the measured values of the reduced Townsend ionization coefficient lie within the sensitivity interval, indicating that a small adjustment ({$<25\%$}) of the proposed cross sections can account for the observed deviations regarding this swarm parameter. 

\section{Production rates for hydrogen radicals}\label{sec:radicals}
\begin{figure*}
\centering
\begin{minipage}[t]{.49\textwidth}
    \includegraphics[width=\textwidth]{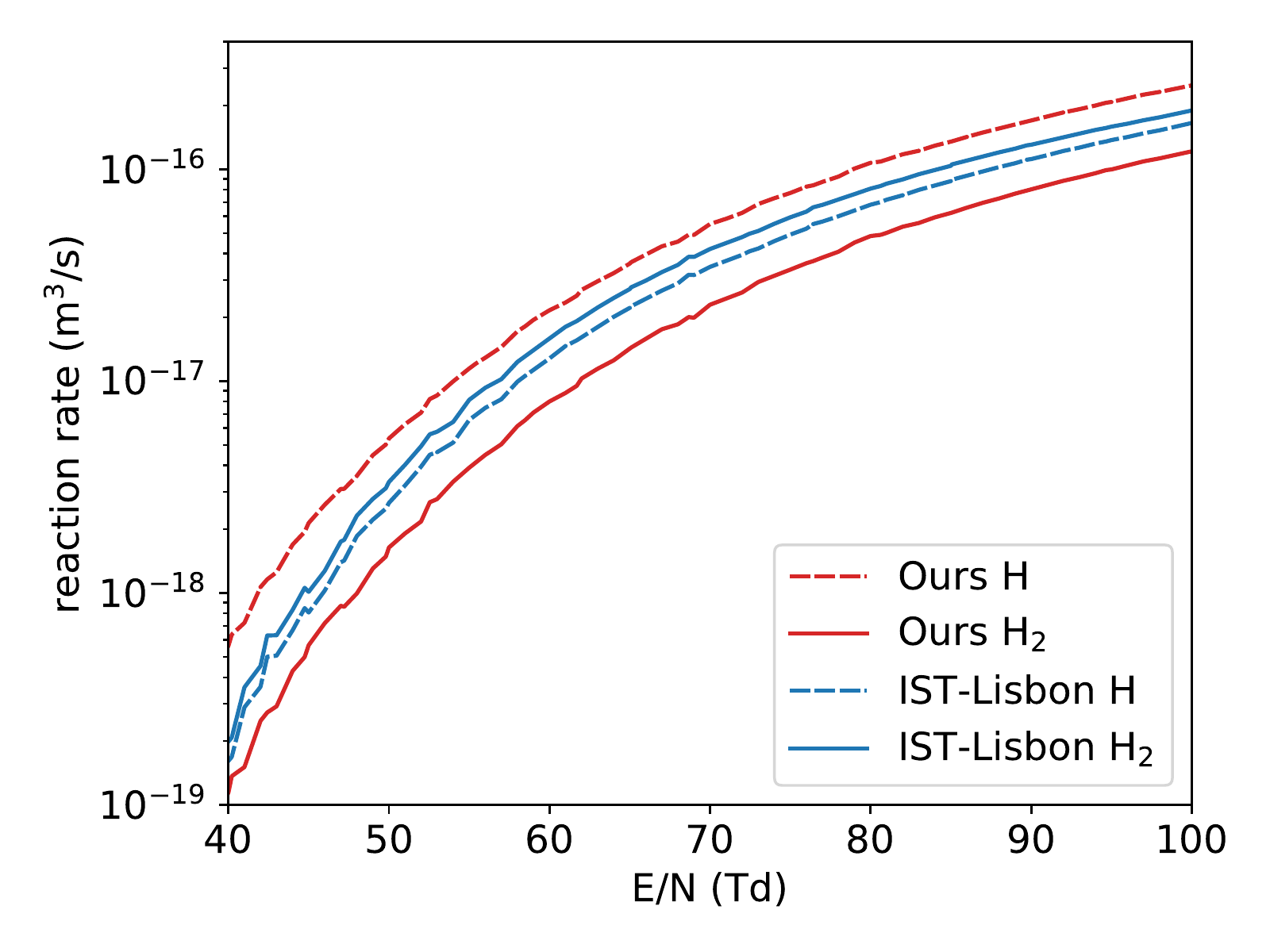}
    \subcaption{Low $E/N$}
    \label{fig:radicals_high}
\end{minipage}\qquad
\begin{minipage}[t]{.49\textwidth}
    \includegraphics[width=\textwidth]{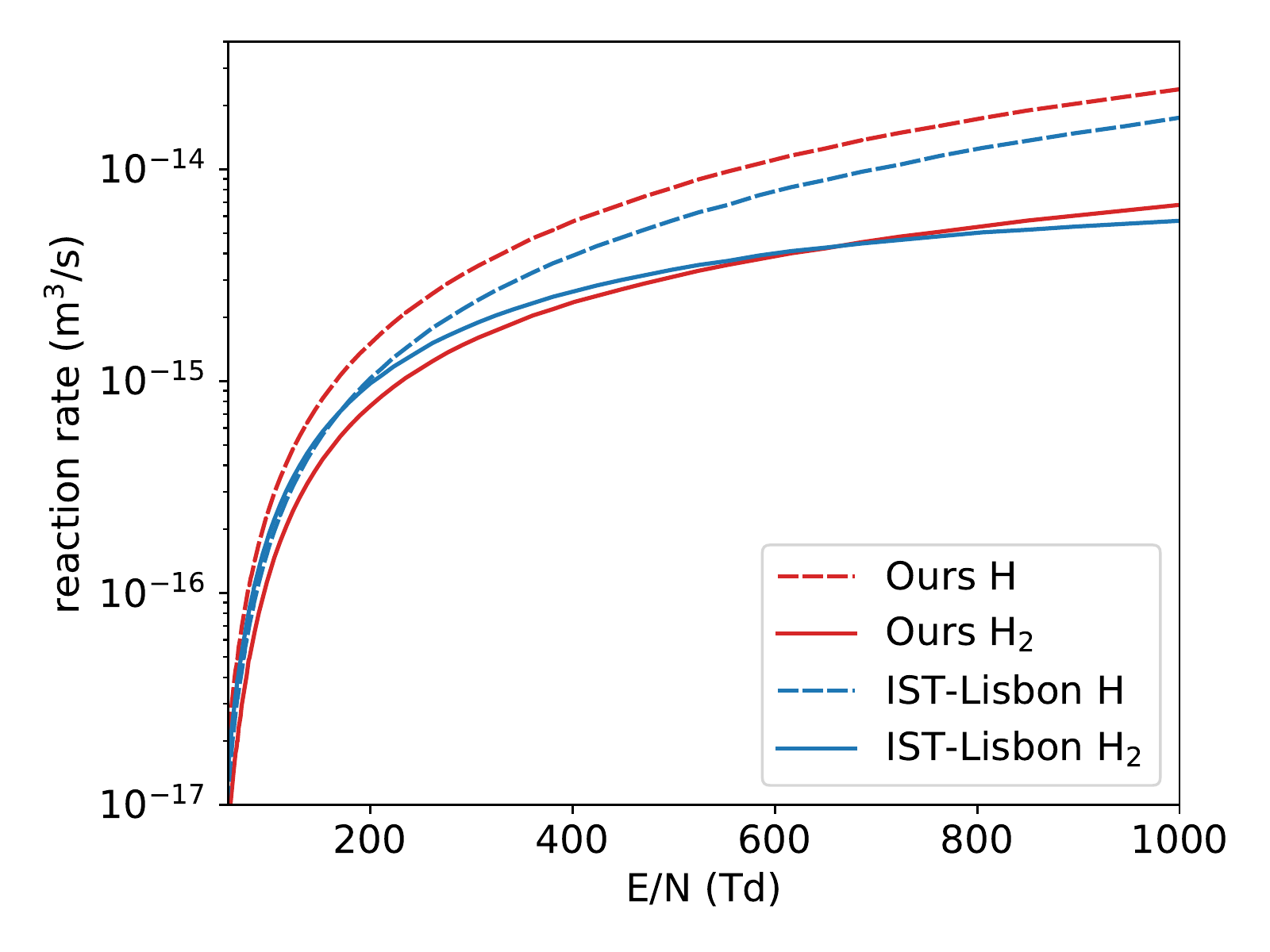}
    \subcaption{High $E/N$}
    \label{fig:radicals_low}
\end{minipage}
    \caption{The predicted reaction rates for the production of the hydrogen species H and H$_2$ for the (swarm-fitted) IST-Lisbon set and Ours (unfitted). Although both cross section sets can be considered consistent (which refers to behaviour of electron swarms only), they exhibit clear differences in the prediction of hydrogen species production.}
    \label{fig:radicals}
\end{figure*}

In the previous section we have introduced two consistent data sets: IST-Lisbon \eqref{it:Lisbon} and Ours \eqref{it:Ours}. The fundamental difference between these two sets is that \eqref{it:Ours} is unfitted and consistent, whereas \eqref{it:Lisbon} employs a fitting procedure to ensure reproduction of swarm parameters. The use of such data-fitting techniques has already been discussed in section \ref{sec:demands}. Here we will illustrate how both data sets predict the production of atomic and molecular hydrogen by inspecting the sum of the reaction rates of hydrogen-producing electron collisions as calculated by 
a Monte-Carlo solver \cite{particle_swarm} based on the modelling framework presented in \cite{Teunissen2016}. The simulations are performed at standard temperature and pressure.

In order to make such a comparison we need ratios regarding the by-products of dissociative electron collisions. However, such data is virtually non-existent. For example: there are no cross sections which distinguish the neutral dissociation processes:
\begin{numcases}{\text{e + \CH{4}{}}\rightarrow}
    \text{e + \CH{2}{} + H}_2 \label{eq:channelCH2_H2}\\ 
    \text{e + \CH{2}{} + 2H} \label{eq:channelCH2_2H}
\end{numcases}
It is known that the dissociation energy of the relatively strong hydrogen bond is $4.52$~eV, therefore it can be expected that due to this additional energy barrier the reaction rate of equation \eqref{eq:channelCH2_2H} will be lower than equation \eqref{eq:channelCH2_H2}. However, without direct observations such arguments will always remain qualitative. For the current purpose of comparing the radical yields of the two data sets, we will assume that the composition of hydrogen products will always be in the lowest energy state. In other words, we assume that reactions like equation \eqref{eq:channelCH2_2H}, which requires additional energy for dissociation, will not occur. The effect is that we will underestimate atomic hydrogen yield, and subsequently overestimate the production of molecular hydrogen.

With this assumption, the reaction rates for hydrogen production have been calculated for both data sets; they are shown in figure~\ref{fig:radicals}. It can be seen that data set \eqref{it:Lisbon} predicts atomic hydrogen yields, approximately $35\%$ lower than \eqref{it:Ours} above $100$\,Td. Similar deviations are also observed for the molecular hydrogen production. For instance, above $100$\,Td the maximum deviation is $45\%$. However, for reduced electric fields below $100$\,Td the deviations between the predictions of production rates for molecular hydrogen are increasing. For instance, at $50$\,Td data set \eqref{it:Lisbon} predicts a molecular hydrogen yield which is $125\%$ higher than data set \eqref{it:Ours}. For atomic hydrogen we find a difference around $50\%$ at $50$\,Td.

These deviations between the production rates of chemical species of two consistent sets clearly illustrate the non-uniqueness of swarm-fitted  data sets. Whether the errors on the production rates for chemical species introduced by relying on data-fitting are tolerable is always dependent on the application and the extent of adjustments performed. However, given the highly reactive nature of atomic hydrogen and the nonlinear nature of plasma-chemical applications, such deviations have to be treated with care.

\section{Summary and Outlook}
\subsection{Summary}
The main contribution of this article are the cross sections for the neutral dissociation of the ground state of \CH{4}{} by electron impact. Secondly, we have used these values to arrive at a complete and consistent cross sections for electron collisions with methane for reduced electric fields between $0.1$\,Td and $1000$\,Td, without relying on any data-fitting techniques. This data set is largely based on the recommendations of Song~\textit{et al.}  \cite{song_cross_2015}, with the addition of a blend of empirical and analytical cross sections for the remaining neutral dissociation processes.

Furthermore this work includes a Boltzmann analysis using a Monte-Carlo solver. We have shown that the presented set of cross sections reproduces measured swarm parameters with maximum deviations of: $35\%$ for ionization, $7.5\%$ for mobility and $20\%$ for characteristic energy.

The presented cross section set distinguishes itself from other data sets by not relying on any data-fitting techniques to ensure consistency. This feature makes our cross section set independent of the limitations imposed by the swarm-fitting procedure. This can be especially attractive for applications that focus on plasma-chemical activation of the gas, such as plasma-assisted vapour deposition, low-temperature methane reforming, etc. Moreover, the absence of any data fitting means that the presented cross section set can be used in a variety of plasma-modelling approaches (e.g.\ hydrodynamic, multi-term Boltzmann or Monte-Carlo/PIC).  

\subsection{Outlook}
The validity of the cross sections proposed in this work has been considered by comparing measured and calculated swarm parameters. In principle this is an implicit metric, since the set of cross sections as a whole is considered as opposed to individual cross sections. However, in section \ref{sec:compare} we have assumed that the recommendations of Song~\textit{et al.} \cite{song_cross_2015} have a sufficiently low error margin such that deviations in the reduced Townsend ionization coefficient can be primarily attributed to the proposed cross sections for neutral dissociation. Although this assumption enables much of the steps taken in this work, it does not give explicit certainty. One way to improve on this is by studying the swarm parameters of mixtures of methane with rare gases \cite{petrovic2007kinetic}. For example, the swarm parameters in Ar-\CH{4}{} mixtures are studied by Sebastian and Wadehra \cite{Sebastian2005}. Still, benchmark experiments for individual cross sections remain highly desirable if the difficulty of diagnosing neutral radical fragments can be overcome.

On the side of computation it would be very desirable to see work in the style of Zio\l kowski~\textit{et al.} \cite{ziolkowski_modeling_2012} (based on R-matrix calculations of electron excitation of methane followed by quasi-classical trajectory simulations with surface hopping) extended to higher electron collision energy than $17$\,eV. The work of Brigg~\textit{et al.}\cite{brigg_r-matrix_2014} highlights some electronic structure issues with these computations and in particular they recommend a multi-reference configuration interaction approach to deal with the multiply-excited target states that are important at high impact energy. (However, the neutral dissociation cross section calculations in Brigg~\textit{et al.} \cite{brigg_r-matrix_2014} are limited to electron impact energies below $15$\,eV and they do not supplant the results from Zio\l kowski~\textit{et al.} \cite{ziolkowski_modeling_2012}). Such R-matrix calculations and trajectory simulations would naturally predict branching ratios between $2$H and H$_2$ channels as well, although an assessment of the importance of zero-point energy (quantized vibrational energy in molecular fragments) should be made. However, there are tools (such as ring polymer Molecular Dynamics \cite{habershon_ring-polymer_2013}) to incorporate this quantum effect into trajectory calculations.

\section*{Acknowledgements}
The authors would like to thank Grzegorz Karwasz and Jonathan Tennyson for their expertise and valuable feedback during correspondence. D.B.\ acknowledges funding through the Dutch TTW-project 16480 ``Making Plasma-Assisted Combustion Efficient''. A.M.\ has received funding from the European Union’s Horizon 2020 research and innovation program under project ID 722337 (SAINT).

\section*{Availability of data}
The data set will be made available on \mbox{\url{www.lxcat.net}} \cite{LxCat}. Furthermore, cross sections obtained in this study are also available in analytical form and as tabulated data in \ref{app:fit} and \ref{app:tabulated}, respectively. The code for the Monte-Carlo Boltzmann solver can be found on \mbox{\url{www.gitlab.com/MD-CWI-NL/particle_swarm}} with the version used identified by the commit hash  \mbox{\url{e04a5644}} made on 25th of May 2021.

\appendix

\section{Fitting functions and parameters for used and reported cross sections}\label{app:fit}

The cross sections for total neutral dissociation, all dissociative ionization processes, and dissociative electron attachment were obtained from fits through the data points reported in tables by Song~\textit{et al.} \cite{song_recommended_2020}. The functions used were those reported in Shirai~\textit{et al.} \cite{shirai_analytic_2002}. The fitting parameters were obtained again for this paper.

\subsection{Basis functions}

Shirai~\textit{et al.} \cite{shirai_analytic_2002} used 3 basis functions from which the fitting functions were created:

\begin{align}
    f_1(x) &= \sigma_0 a_1 \left( \frac{x}{\epsilon_R} \right)^{a_2}, \\
    f_2(x) &= \frac{f_1(x)}{\left[ 1 + \left( \frac{x}{a_3} \right)^{a_2 + a_4}\right]}, \\
    f_3(x) &= \frac{f_1(x)}{\left[ 1 + \left( \frac{x}{a_3} \right)^{a_2 + a_4} + \left( \frac{x}{a_5} \right)^{a_2 + a_6} \right]},
\end{align}
with $\sigma_0 = 1 \cdot 10^{-20}$\,m$^2$, $\epsilon_R = 1.361 \cdot 10^{-2}$\,keV (Rydberg constant), and $a_i$ the parameters which will be obtained for each specific reaction by fitting the data points.

\subsection{Dissociative Ionization}

\begin{figure}
    \centering
    \includegraphics[scale=0.4]{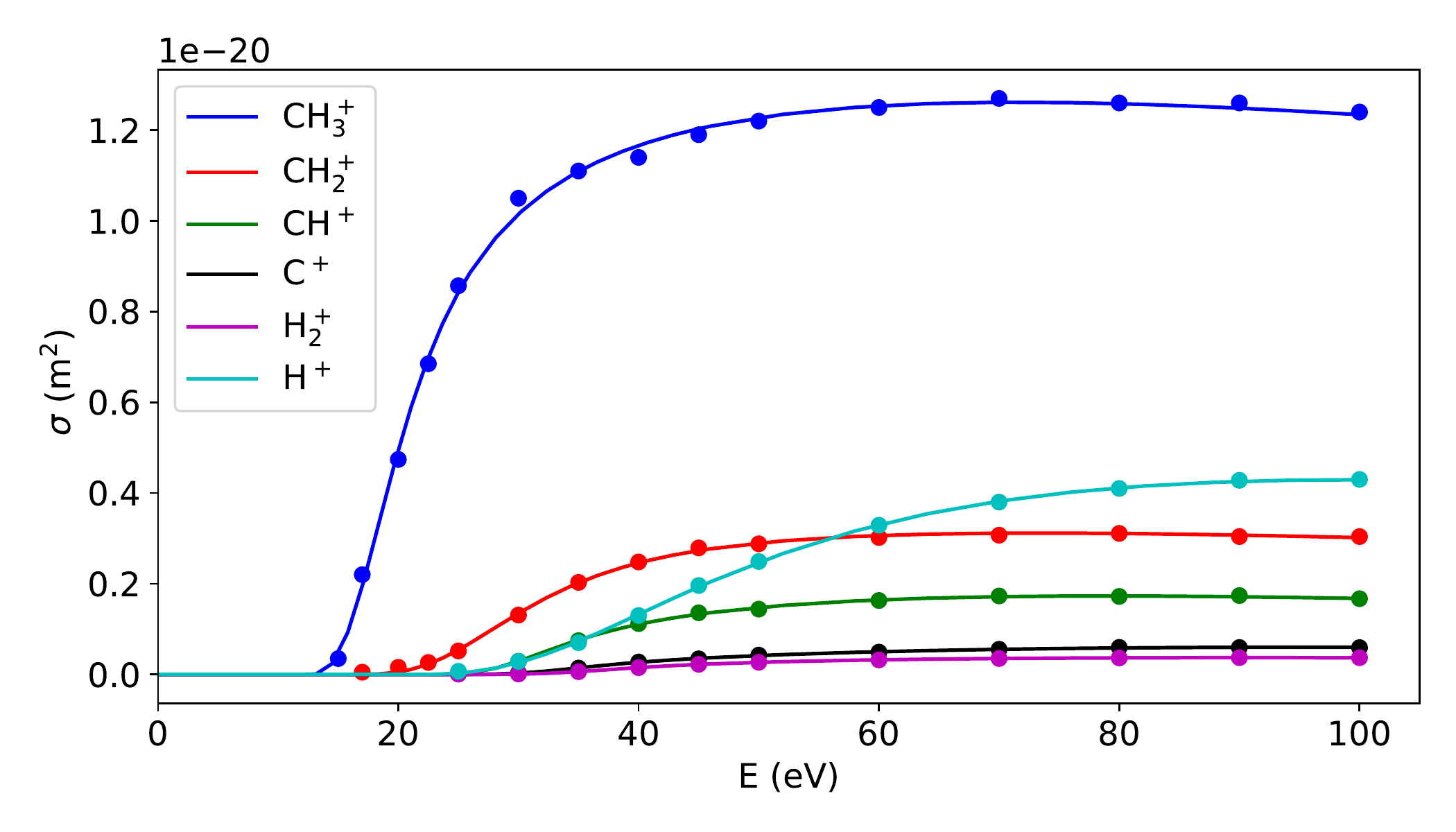}
    \caption{The cross sections of the dissociative ionization reactions of \CH{4}{}. The solid lines are the result of fitting equation \eqref{eq:fit_dissioni} to the tabulated cross sections for these reactions from Song~\textit{et al.} \cite{song_recommended_2020} which are represented by filled circles in the same color.}
    \label{fig:fit_dissioni}
\end{figure}

The fitting function reported by Shirai~\textit{et al.} \cite{shirai_analytic_2002} to be used for the dissociative ionization reactions is the following:

\begin{equation}\label{eq:fit_dissioni}
    \sigma_i(\epsilon) = f_3(\epsilon_1),
\end{equation}
with $\epsilon$ the incident electron energy in keV, $\epsilon_1 = \epsilon - \epsilon_{th}$, and $\epsilon_{th}$ the threshold energy of the reaction in keV. Equation \ref{eq:fit_dissioni} was fitted through the tabulated cross sections and threshold energies for the dissociative ionization reactions reported in Song~\textit{et al.} \cite{song_recommended_2020}. The used data points and resulting fits are shown up to 100\,eV in figure \ref{fig:fit_dissioni}. The fitting parameters are tabulated in table \ref{tab:fit_dissioni}

\begin{table*}
    \centering
    \begin{tabular}{r||c|c|c|c|c|c|c}
          & $\epsilon_{th}$ (eV) & $a_1$ & $a_2$ & $a_3$ & $a_4$ & $a_5$ & $a_6$ \\
         \hline
        \CH{3}{+} & 12.63 & 5.5333 & 2.7119 & 0.0071 & -0.2619 & 0.0194 & 0.8917 \\
        \CH{2}{+} & 16.20 & 0.2575 & 2.9997 & 0.0141 & -0.2828 & 0.0289 & 1.0172 \\
        \CH{}{+} & 22.20 & 0.295 & 3.4235 & 0.0207 & 0.9925 & 0.0100 & -0.5789 \\
        C$^+$ & 22.00 & 0.0392 & 4.6413 & 0.0243 & 1.1558 & 0.0125 & -0.7372 \\
        H$_2^+$ & 22.30 & 0.0134 & 5.0600 & 0.0147 & -0.7746 & 0.0242 & 1.0240 \\
        H$^+$ & 21.10 & 0.0985 & 2.7831 & 0.0210 & -0.6691 & 0.0403 & 1.0503 \\
    \end{tabular}
    \caption{Parameters obtained by fitting equation \eqref{eq:fit_dissioni} to the tabulated cross sections of Song~\textit{et al.} \cite{song_recommended_2020} for dissociative ionization reactions}
    \label{tab:fit_dissioni}
\end{table*}

\subsection{Dissociative Electron Attachment}

\begin{figure}
    \centering
    \includegraphics[scale=0.4]{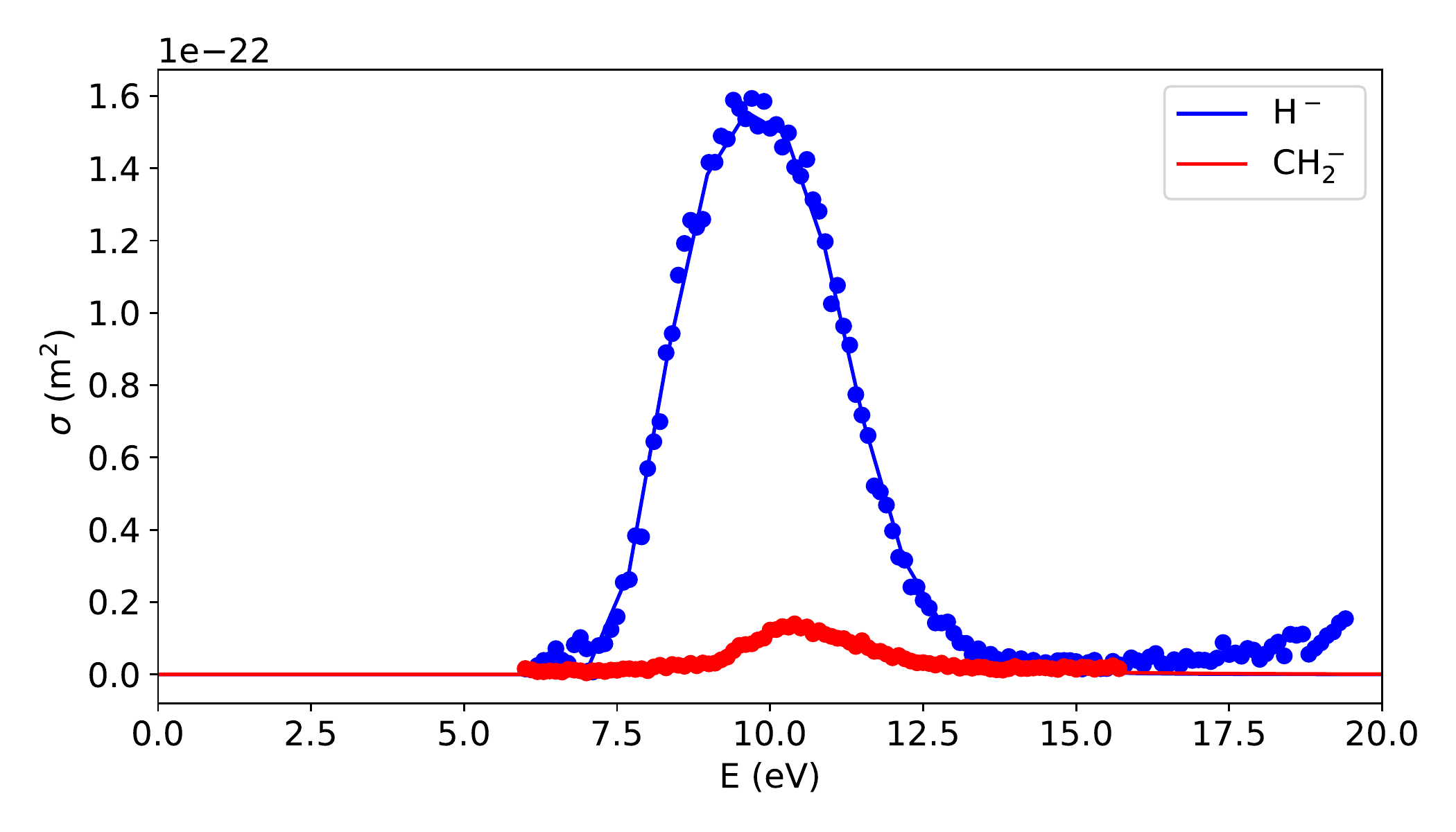}
    \caption{The cross sections of the dissociative electron attachment reactions of \CH{4}{}. The solid lines are the result of fitting equation \eqref{eq:fit_dissioni} to the tabulated cross sections for these reactions from Song~\textit{et al.} \cite{song_recommended_2020} which are represented as filled circles in the same color.}
    \label{fig:fit_dissatt}
\end{figure}

Shirai~\textit{et al.} \cite{shirai_analytic_2002} use the same fitting function for dissociative ionization, equation \eqref{eq:fit_dissioni}, as for fitting the dissociative electron attachment cross sections (including the same definition for $\epsilon_1$). We use this fitting function to fit the tabulated cross sections for dissociative electron attachment reactions from Song~\textit{et al.} \cite{song_recommended_2020}. The fits and corresponding data points are shown in figure \ref{fig:fit_dissatt} and the fitting parameters are reported in table \ref{tab:fit_dissatt}.

\begin{table*}
    \centering
    \begin{tabular}{r||c|c|c|c|c|c|c}
          & $\epsilon_{th}$ (eV) & $a_1$ & $a_2$ & $a_3$ & $a_4$ & $a_5$ & $a_6$ \\
         \hline
        
        H$^-$ & 6.0 & 128.0817 & 5.0736 & 0.0024 & 0.1908 & 0.0041 & 10.1747\\
        \CH{2}{-} & 6.0 & 1.5496 & 3.1405$\cdot 10^{-5}$ & 0.0012 & 4.8957 & 0.0164 & -4.8826\\
    \end{tabular}
    \caption{Parameters obtained by fitting equation \eqref{eq:fit_dissioni} to the tabulated cross sections of Song~\textit{et al.} \cite{song_recommended_2020} for dissociative electron attachment}
    \label{tab:fit_dissatt}
\end{table*}

\subsection{Total Dissociation}

\begin{figure}
    \centering
    \includegraphics[scale=0.4]{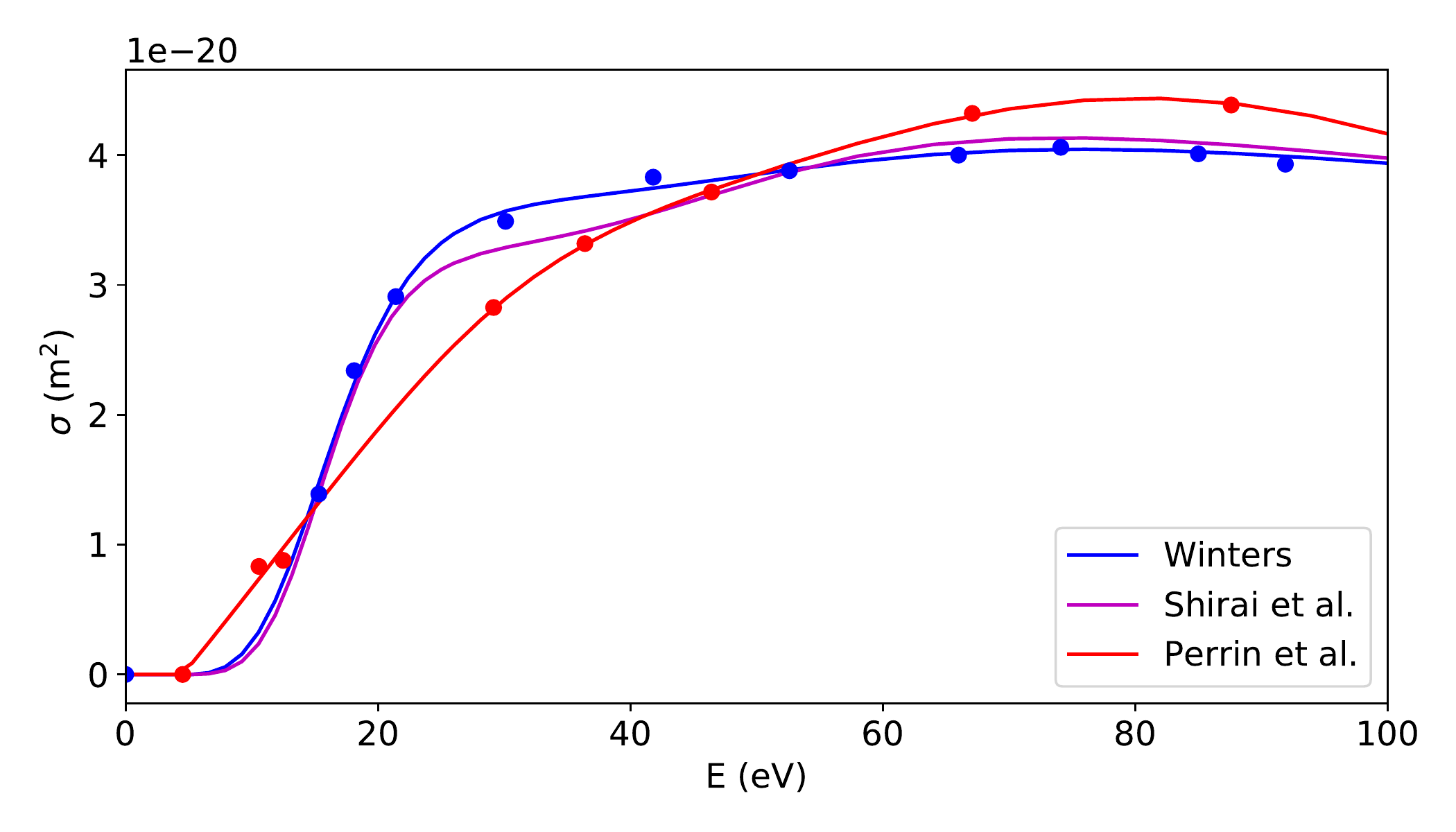}
    \caption{The total dissociation cross sections as measured by Winters \cite{winters_dissociation_1975} (filled blue circles), and by Perrin~\textit{et al.} \cite{perrin_dissociation_1982} (filled red circles). The solid lines in the same colors as the measurement points are the fits done in this paper using equation \eqref{eq:fit_totaldiss}. The solid line of Shirai~\textit{et al.} \cite{shirai_analytic_2002} is obtained by using their reported fitting parameters with equation \eqref{eq:fit_totaldiss}.}
    \label{fig:fit_totaldiss}
\end{figure}

The fitting function for the total dissociation used by Shirai~\textit{et al.} \cite{shirai_analytic_2002} is given by:

\begin{equation}\label{eq:fit_totaldiss}
    \sigma_{\text{TD}}(\epsilon) = f_2(\epsilon_1) + a_5\cdot f_2\left(\frac{\epsilon_1}{a_6}\right),
\end{equation}
where $\epsilon_1$ again has the same definition as for equation \eqref{eq:fit_dissioni}. The total dissociation cross section was measured by Winters \cite{winters_dissociation_1975} and Perrin~\textit{et al.} \cite{perrin_dissociation_1982}. We have obtained data points for both measurements by extracting them from the published graphs using WebPlotDigitizer \cite{Rohatgi2020}. The fits and the data points for both measurements as well as equation \eqref{eq:fit_totaldiss} using the fitting parameters reported by Shirai~\textit{et al.} \cite{shirai_analytic_2002} for total dissociation are shown in figure \ref{fig:fit_totaldiss}. Deviations up to 20\% can arise due to different fitting parameters and fitted data points. These deviations in the total dissociation will propagate to the cross sections of the individual neutral dissociation reactions. Increasing the cross section of the neutral dissociation cross sections has the effect of reducing the Townsend ionization coefficient $\alpha$. In this paper we have used the data points of Winters \cite{winters_dissociation_1975} and the fitting parameters as reported in table \ref{tab:fit_totaldiss}.

\begin{table}
    \centering
    \begin{tabular}{|c|c|}
        \hline
        $\epsilon_{th}$ (eV) & 4.51 \\
        \hline
        $a_1$ & 4.1200 \\
        \hline
        $a_2$ & 3.0594\\
        \hline
        $a_3$ & 0.0142\\
        \hline
        $a_4$ & 0.3606\\
        \hline
        $a_5$ & 0.4630\\
        \hline
        $a_6$ & 3.8830\\
        \hline
    \end{tabular}
    \caption{Parameters obtained by fitting equation \eqref{eq:fit_totaldiss} to the measured cross sections of Winters \cite{winters_dissociation_1975} for total dissociation.}
    \label{tab:fit_totaldiss}
\end{table}

\subsection{Neutral Dissociation to \CH{2}{}}

\begin{figure}
    \centering
    \includegraphics[scale=0.4]{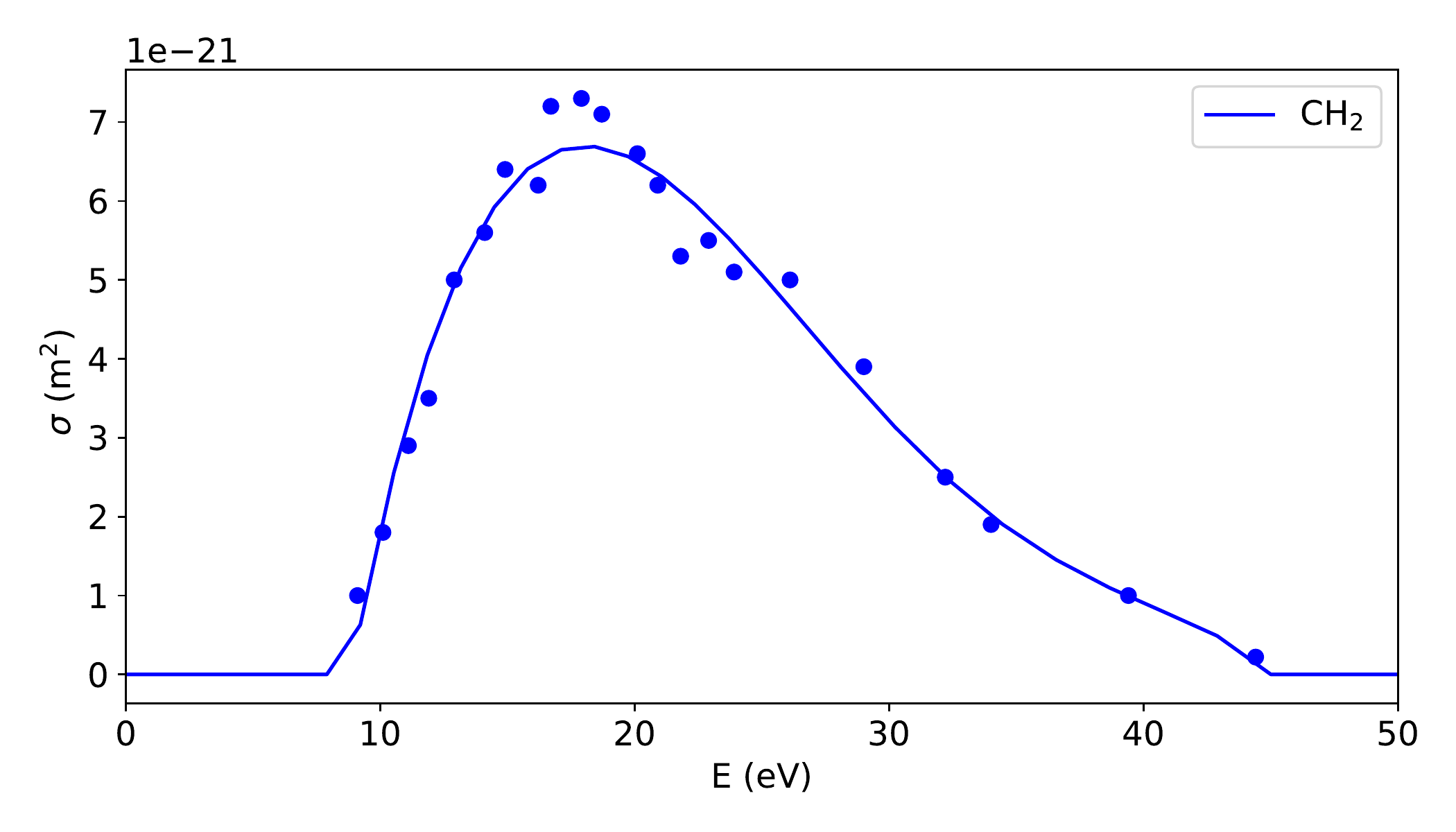}
    \caption{The cross sections for neutral dissociation to \CH{2}{} as measured by Nakano~\textit{et al.} \cite{nakano_electron-impact_1991-1, nakano_electron-impact_1991} and the corresponding fit obtained by equation \eqref{eq:fit_neutraldiss_ch2} combined with the fitting parameters in table \ref{tab:fit_neutraldiss_ch2}}
    \label{fig:fit_neutraldiss_ch2}
\end{figure}

In this paper we have taken the measured cross sections for neutral dissociation into \CH{2}{} from Nakano~\textit{et al.} \cite{nakano_electron-impact_1991-1, nakano_electron-impact_1991}. To smooth the data we have fitted a fourth order polynomial through the data points:

\begin{equation}\label{eq:fit_neutraldiss_ch2}
    f(\epsilon) = a_0 + a_1 \epsilon + a_2 \epsilon^2 + a_3 \epsilon^3 + a_4 \epsilon^4,
\end{equation}
with fitting parameters $a_i$, and $\epsilon$ the incident electron energy in eV. Note that this function is only valid within the bounds of the measurement energies i.e.\ \mbox{$9.1$\,eV $\leq \epsilon \leq 44.4$\,eV}. The fit and the corresponding data points are shown in figure \ref{fig:fit_neutraldiss_ch2}. The fitting parameters are reported in table \ref{tab:fit_neutraldiss_ch2}.

\begin{table}
    \centering
    \begin{tabular}{|c|r|}
        \hline
        $a_0$ & $-3.0203\cdot 10^{-20}$\\
        \hline
        $a_1$ & $5.4772\cdot 10^{-21}$ \\
        \hline
        $a_2$ & $-2.8119\cdot 10^{-22}$ \\
        \hline
        $a_3$ & $5.8213\cdot 10^{-24}$ \\
        \hline
        $a_4$ & $-4.3221\cdot 10^{-26}$ \\
        \hline
    \end{tabular}
    \caption{Parameters obtained by fitting equation \eqref{eq:fit_neutraldiss_ch2} to the measured cross sections of Nakano~\textit{et al.}\cite{nakano_electron-impact_1991-1, nakano_electron-impact_1991} for neutral dissociation to \CH{2}{}}
    \label{tab:fit_neutraldiss_ch2}
\end{table}

\section{Tabulated cross sections for neutral dissociation to \CH{3}{}, \CH{}{}, and C}\label{app:tabulated}

Calculated cross sections for neutral dissociation into \CH{3}{}, \CH{}{} and C are reported in table \ref{tab:cs_neutraldiss_ch3}, \ref{tab:cs_neutraldiss_ch}, and \ref{tab:cs_neutraldiss_c}, respectively.

\begin{table}
    \centering
    \begin{tabular}{|c|c|}
    \hline
    $\epsilon$ (eV) & $\sigma_{\CH{}{}}$ (m$^2$) \\
    \hline
 15.79 &	$2.2169\cdot 10^{-29}$ \\
 \hline
 17.11 &	$3.4360\cdot 10^{-28}$\\
 \hline
 18.42 &	$8.9545\cdot 10^{-28}$\\
 \hline
 19.74 &	$1.5509\cdot 10^{-27}$\\
 \hline
 21.05 &	$2.2067\cdot 10^{-27}$\\
 \hline
 22.37 &	$2.8625\cdot 10^{-27}$\\
 \hline
 23.68 &	$4.3996\cdot 10^{-24}$\\
 \hline
 25.00 &	$3.4406\cdot 10^{-23}$\\
 \hline
 26.00 &	$8.8330\cdot 10^{-23}$\\
 \hline
 28.11 &	$3.0330\cdot 10^{-22}$\\
 \hline
 30.22 &	$5.9931\cdot 10^{-22}$\\
 \hline
 32.33 &	$8.7592\cdot 10^{-22}$\\
 \hline
 34.44 &	$1.0797\cdot 10^{-21}$\\
 \hline
 36.56 &	$1.2107\cdot 10^{-21}$\\
 \hline
 38.67 &	$1.2892\cdot 10^{-21}$\\
 \hline
 40.78 &	$1.3348\cdot 10^{-21}$\\
 \hline
 42.89 &	$1.3612\cdot 10^{-21}$\\
 \hline
 45.00 &	$1.3769\cdot 10^{-21}$\\
 \hline
 46.00 &	$1.3820\cdot 10^{-21}$\\
 \hline
 52.00 &	$1.3999\cdot 10^{-21}$\\
 \hline
 58.00 &	$1.4092\cdot 10^{-21}$\\
 \hline
 64.00 &	$1.4120\cdot 10^{-21}$\\
 \hline
 70.00 &	$1.4063\cdot 10^{-21}$\\
 \hline
 76.00 &	$1.3920\cdot 10^{-21}$\\
 \hline
 82.00 &	$1.3706\cdot 10^{-21}$\\
 \hline
 88.00 &	$1.3441\cdot 10^{-21}$\\
 \hline
 94.00 &	$1.3146\cdot 10^{-21}$\\
 \hline
 100.0 &	$1.2839\cdot 10^{-21}$\\
    \hline
    \end{tabular}
    \caption{Calculated cross sections for neutral dissociation to \CH{}{}. Threshold energy is 15.5\,eV.}
    \label{tab:cs_neutraldiss_ch}
\end{table}

\begin{table}
    \centering
    \begin{tabular}{|c|c|}
    \hline
         $\epsilon$ (eV) & $\sigma_{\CH{3}{}}$ (m$^2$) \\
         \hline
 7.90 &   $1.7980\cdot 10^{-22}$ \\
          \hline
 9.21 &	$1.6220\cdot 10^{-21}$ \\ 
 \hline
 10.53 &	$3.8168\cdot 10^{-21}$ \\
 \hline
 11.84 &	$6.2953\cdot 10^{-21}$ \\
 \hline
 13.16 &	$8.7738\cdot 10^{-21}$ \\
 \hline
 14.47 &	$1.2135\cdot 10^{-20}$ \\
 \hline
 15.79 &	$1.5230\cdot 10^{-20}$ \\ 
 \hline
 17.11 &	$1.7797\cdot 10^{-20}$ \\
 \hline
 18.42 &	$1.9773\cdot 10^{-20}$ \\ 
 \hline
 19.74 &	$2.1205\cdot 10^{-20}$ \\ 
 \hline
 21.05 &	$2.2160\cdot 10^{-20}$ \\
 \hline
 22.37 &	$2.2697\cdot 10^{-20}$ \\
 \hline
 23.68 &	$2.2872\cdot 10^{-20}$ \\
 \hline
 25.00 &	$2.2732\cdot 10^{-20}$ \\
 \hline
 26.00 &	$2.2445\cdot 10^{-20}$ \\
 \hline
 28.11 &	$2.1454\cdot 10^{-20}$ \\
 \hline
 30.22 &	$2.0177\cdot 10^{-20}$ \\
 \hline
 32.33 &	$1.8871\cdot 10^{-20}$ \\
 \hline
 34.44 &	$1.7693\cdot 10^{-20}$ \\ 
 \hline
 36.56 &	$1.6706\cdot 10^{-20}$ \\
 \hline
 38.67 &	$1.5914\cdot 10^{-20}$ \\
 \hline
 40.78 &	$1.5296\cdot 10^{-20}$ \\
 \hline
 42.89 &	$1.4821\cdot 10^{-20}$ \\
 \hline
 45.00 &	$1.4462\cdot 10^{-20}$ \\
 \hline
 46.00 &	$1.4324\cdot 10^{-20}$ \\
 \hline
 52.00 &	$1.3817\cdot 10^{-20}$ \\
 \hline
 58.00 &	$1.3624\cdot 10^{-20}$ \\
 \hline
 64.00 &	$1.3547\cdot 10^{-20}$ \\
 \hline
 70.00 &	$1.3482\cdot 10^{-20}$ \\
 \hline
 76.00 &	$1.3388\cdot 10^{-20}$ \\
 \hline
 82.00 &	$1.3259\cdot 10^{-20}$ \\ 
 \hline
 88.00 &	$1.3102\cdot 10^{-20}$ \\
 \hline
 94.00 &	$1.2926\cdot 10^{-20}$ \\
 \hline
 100.0 &	$1.2743\cdot 10^{-20}$ \\
 \hline
    \end{tabular}
    \caption{Calculated cross sections for neutral dissociation to \CH{3}{}. Threshold energy is 7.5\,eV.}
    \label{tab:cs_neutraldiss_ch3}
\end{table}

\begin{table}
    \centering
    \begin{tabular}{|c|c|}
    \hline
    $\epsilon$ (eV) & $\sigma_\text{C}$ (m$^2$) \\
    \hline
 15.79 &	$5.2967\cdot 10^{-31}$\\
    \hline
 17.11 &	$8.2093\cdot 10^{-30}$\\
    \hline
 18.42 &	$2.1394\cdot 10^{-29}$\\
    \hline
 19.74 &	$3.7055\cdot 10^{-29}$\\
    \hline
 21.05 &	$5.2723\cdot 10^{-29}$\\
    \hline
 22.37 &	$6.8390\cdot 10^{-29}$\\
    \hline
 23.68 &	$7.1010\cdot 10^{-26}$\\
    \hline
 25.00 &	$9.3899\cdot 10^{-25}$\\
    \hline
 26.00 &	$3.3122\cdot 10^{-24}$\\
    \hline
 28.11 &	$1.9703\cdot 10^{-23}$\\
    \hline
 30.22 &	$6.0694\cdot 10^{-23}$\\
    \hline
 32.33 &	$1.2475\cdot 10^{-22}$\\
    \hline
 34.44 &	$1.9488\cdot 10^{-22}$\\
    \hline
 36.56 &	$2.5467\cdot 10^{-22}$\\
    \hline
 38.67 &	$2.9881\cdot 10^{-22}$\\
    \hline
 40.78 &	$3.2963\cdot 10^{-22}$\\
    \hline
 42.89 &	$3.5130\cdot 10^{-22}$\\
    \hline
 45.00 & $3.6723\cdot 10^{-22}$\\
    \hline
 46.00 &	$3.7348\cdot 10^{-22}$\\
    \hline
 52.00 &	$4.0210\cdot 10^{-22}$\\
    \hline
 58.00 &	$4.2381\cdot 10^{-22}$\\
    \hline
 64.00 &	$4.4125\cdot 10^{-22}$\\
    \hline
 70.00 &	$4.5416\cdot 10^{-22}$\\
    \hline
 76.00 &	$4.6243\cdot 10^{-22}$\\
    \hline
 82.00 &	$4.6645\cdot 10^{-22}$\\
    \hline
 88.00 &	$4.6697\cdot 10^{-22}$\\
    \hline
 94.00 &	$4.6478\cdot 10^{-22}$\\
    \hline
 100.0 &	$4.6063\cdot 10^{-22}$\\
    \hline
    \end{tabular}
    \caption{Calculated cross sections for neutral dissociation to C. Threshold energy is 15.5\,eV.}
    \label{tab:cs_neutraldiss_c}
\end{table}

\clearpage

\section*{References}
\bibliography{References}

\providecommand{\newblock}{}
\begin{thebibliography}{10}
\expandafter\ifx\csname url\endcsname\relax
  \def\url#1{{\tt #1}}\fi
\expandafter\ifx\csname urlprefix\endcsname\relax\def\urlprefix{URL }\fi
\providecommand{\eprint}[2][]{\url{#2}}

\bibitem{starikovskaia_plasma_2006}
Starikovskaia S~M 2006 {\em J. Phys. D: Appl. Phys.\/} {\bf 39} R265--R299 ISSN
  0022-3727 publisher: IOP Publishing
  \urlprefix\url{https://doi.org/10.1088/0022-3727/39/16/r01}

\bibitem{starikovskiy_plasma-assisted_2013}
Starikovskiy A and Aleksandrov N 2013 {\em Progress in Energy and Combustion
  Science\/} {\bf 39} 61--110 ISSN 0360-1285
  \urlprefix\url{https://www.sciencedirect.com/science/article/pii/S0360128512000354}

\bibitem{nozaki_non-thermal_2013}
Nozaki T and Okazaki K 2013 {\em Catalysis Today\/} {\bf 211} 29--38 ISSN
  0920-5861
  \urlprefix\url{https://www.sciencedirect.com/science/article/pii/S092058611300148X}

\bibitem{kamo_diamond_1983}
Kamo M, Sato Y, Matsumoto S and Setaka N 1983 {\em Journal of Crystal Growth\/}
  {\bf 62} 642--644 ISSN 0022-0248
  \urlprefix\url{https://www.sciencedirect.com/science/article/pii/0022024883904116}

\bibitem{yair_study_2009}
Yair Y, Takahashi Y, Yaniv R, Ebert U and Goto Y 2009 {\em Journal of
  Geophysical Research: Planets\/} {\bf 114} ISSN 2156-2202
  \urlprefix\url{https://agupubs.onlinelibrary.wiley.com/doi/abs/10.1029/2008JE003311}

\bibitem{Kohn2019}
K{\"{o}}hn C, Dujko S, Chanrion O and Neubert T 2019 {\em Icarus\/} {\bf 333}
  294--305 ISSN 10902643 (\textit{Preprint} \eprint{1802.09906})

\bibitem{horton_atomic_1996}
Horton L~D 1996 {\em Phys. Scr.\/} {\bf T65} 175--178 ISSN 1402-4896 publisher:
  IOP Publishing \urlprefix\url{https://doi.org/10.1088/0031-8949/1996/t65/025}

\bibitem{pitchford_comparisons_2013}
Pitchford L~C, Alves L~L, Bartschat K, Biagi S~F, Bordage M~C, Phelps A~V,
  Ferreira C~M, Hagelaar G~J~M, Morgan W~L, Pancheshnyi S, Puech V, Stauffer A
  and Zatsarinny O 2013 {\em J. Phys. D: Appl. Phys.\/} {\bf 46} 334001 ISSN
  0022-3727 publisher: IOP Publishing
  \urlprefix\url{https://doi.org/10.1088/0022-3727/46/33/334001}

\bibitem{buckman_atomic_2013}
Buckman S~J, Brunger M~J and Ratnavelu K 2013 {\em Fusion Science and
  Technology\/} {\bf 63} 385--391 ISSN 1536-1055 publisher: Taylor \& Francis
  \urlprefix\url{https://doi.org/10.13182/FST13-A16446}

\bibitem{petrovic2007kinetic}
Petrovi{\'{c}} Z~L, {\v{S}}uvakov M, Nikitovi{\'{c}} Z, Dujko S,
  {\v{S}}ai{\'{c}} O, Jovanovi{\'{c}} J, Malovi{\'{c}} G and Stojanovi{\'{c}} V
  2007 {\em Plasma Sources Science and Technology\/} {\bf 16} ISSN 09630252

\bibitem{alves_ist-lisbon_2014}
Alves L~L 2014 {\em J. Phys.: Conf. Ser.\/} {\bf 565} 012007 ISSN 1742-6596
  publisher: IOP Publishing
  \urlprefix\url{https://doi.org/10.1088/1742-6596/565/1/012007}

\bibitem{Crompton1994}
Crompton R 1994 {\em Benchmark measurements of cross sections for electron
  collisions: Analysis of scattered electrons\/} vol~33 ISBN 0120038331

\bibitem{song_recommended_2020}
Song M~Y, Yoon J~S, Cho H, Karwasz G~P, Kokoouline V, Nakamura Y and Tennyson J
  2020 {\em Eur. Phys. J. D\/} {\bf 74} 60 ISSN 1434-6079
  \urlprefix\url{https://doi.org/10.1140/epjd/e2020-100543-6}

\bibitem{song_cross_2015}
Song M~Y, Yoon J~S, Cho H, Itikawa Y, Karwasz G~P, Kokoouline V, Nakamura Y and
  Tennyson J 2015 {\em Journal of Physical and Chemical Reference Data\/} {\bf
  44} 023101 ISSN 0047-2689 publisher: American Institute of Physics
  \urlprefix\url{https://aip.scitation.org/doi/10.1063/1.4918630}

\bibitem{allan_improved_2007}
Allan M 2007 {\em AIP Conference Proceedings\/} {\bf 901} 107--116 ISSN
  0094-243X publisher: American Institute of Physics
  \urlprefix\url{https://aip.scitation.org/doi/abs/10.1063/1.2727361}

\bibitem{fedus_ramsauer-townsend_2014}
Fedus K and Karwasz G~P 2014 {\em Eur. Phys. J. D\/} {\bf 68} 93 ISSN 1434-6079
  \urlprefix\url{https://doi.org/10.1140/epjd/e2014-40738-x}

\bibitem{sakae_scattering_1989}
Sakae T, Sumiyoshi S, Murakami E, Matsumoto Y, Ishibashi K and Katase A 1989
  {\em J. Phys. B: At. Mol. Opt. Phys.\/} {\bf 22} 1385--1394 ISSN 0953-4075
  publisher: IOP Publishing
  \urlprefix\url{https://doi.org/10.1088/0953-4075/22/9/011}

\bibitem{itikawa_63_2003}
Elford M~T, Buckman S~J and Brunger M 2003 6.3 {Elastic} momentum transfer
  cross sections {\em Interactions of {Photons} and {Electrons} with
  {Molecules}\/} vol 17C ed Itikawa Y (Berlin/Heidelberg: Springer-Verlag) pp
  6085--6117 ISBN 978-3-540-44338-4 series Title: Landolt-Börnstein - Group I
  Elementary Particles, Nuclei and Atoms
  \urlprefix\url{http://materials.springer.com/lb/docs/sm_lbs_978-3-540-45843-2_6}

\bibitem{Kurachi_1990}
Kurachi M and Nakamura Y 1990 {\em Proceedings of 13th Symposium on Ion Sources
  and Ion-Assisted Technology\/}

\bibitem{itikawa_51_2003}
Lindsay B~G and Mangan M~A 2003 5.1 {Ionization} {\em Interactions of {Photons}
  and {Electrons} with {Molecules}\/} vol 17C ed Itikawa Y (Berlin/Heidelberg:
  Springer-Verlag) pp 5001--5077 ISBN 978-3-540-44338-4 series Title:
  Landolt-Börnstein - Group I Elementary Particles, Nuclei and Atoms
  \urlprefix\url{http://materials.springer.com/lb/docs/sm_lbs_978-3-540-45843-2_2}

\bibitem{rawat_absolute_2008}
Rawat P, Prabhudesai V~S, Rahman M~A, Ram N~B and Krishnakumar E 2008 {\em
  International Journal of Mass Spectrometry\/} {\bf 277} 96--102 ISSN
  1387-3806
  \urlprefix\url{https://www.sciencedirect.com/science/article/pii/S1387380608001917}

\bibitem{gadoum_set_2019}
Gadoum A and Benyoucef D 2019 {\em IEEE Transactions on Plasma Science\/} {\bf
  47} 1505--1513 ISSN 1939-9375 conference Name: IEEE Transactions on Plasma
  Science

\bibitem{erwin_electron-impact_2005}
Erwin D~A and Kunc J~A 2005 {\em Phys. Rev. A\/} {\bf 72} 052719 publisher:
  American Physical Society
  \urlprefix\url{https://link.aps.org/doi/10.1103/PhysRevA.72.052719}

\bibitem{erwin_dissociation_2008}
Erwin D~A and Kunc J~A 2008 {\em Journal of Applied Physics\/} {\bf 103} 064906
  ISSN 0021-8979 publisher: American Institute of Physics
  \urlprefix\url{https://aip.scitation.org/doi/10.1063/1.2891694}

\bibitem{motlagh_cross_1998}
Motlagh S and Moore J~H 1998 {\em J. Chem. Phys.\/} {\bf 109} 432--438 ISSN
  0021-9606 publisher: American Institute of Physics
  \urlprefix\url{https://aip.scitation.org/doi/10.1063/1.476580}

\bibitem{janev_collision_2002}
Janev R~K and Reiter D 2002 {\em Physics of Plasmas\/} {\bf 9} 4071--4081 ISSN
  1070-664X publisher: American Institute of Physics
  \urlprefix\url{https://aip.scitation.org/doi/10.1063/1.1500735}

\bibitem{reiter_hydrocarbon_2010}
Reiter D and Janev R~K 2010 {\em Contributions to Plasma Physics\/} {\bf 50}
  986--1013 ISSN 1521-3986
  \urlprefix\url{https://onlinelibrary.wiley.com/doi/abs/10.1002/ctpp.201000090}

\bibitem{ziolkowski_modeling_2012}
Zi{\'{o}}{\l}kowski M, Vik{\'{a}}r A, Mayes M~L, Bencsura {\'{A}}, Lendvay G
  and Schatz G~C 2012 {\em J. Chem. Phys.\/} {\bf 137} 22A510 ISSN 0021-9606
  publisher: American Institute of Physics
  \urlprefix\url{https://aip.scitation.org/doi/10.1063/1.4733706}

\bibitem{nakano_electron-impact_1991-1}
Nakano T, Toyoda H~T~H and Sugai H~S~H 1991 {\em Jpn. J. Appl. Phys.\/} {\bf
  30} 2912 ISSN 1347-4065 publisher: IOP Publishing
  \urlprefix\url{https://iopscience.iop.org/article/10.1143/JJAP.30.2912/meta}

\bibitem{nakano_electron-impact_1991}
Nakano T, Toyoda H~T~H and Sugai H~S~H 1991 {\em Jpn. J. Appl. Phys.\/} {\bf
  30} 2908 ISSN 1347-4065 publisher: IOP Publishing
  \urlprefix\url{https://iopscience.iop.org/article/10.1143/JJAP.30.2908/meta}

\bibitem{makochekanwa_experimental_2006}
Makochekanwa C, Oguri K, Suzuki R, Ishihara T, Hoshino M, Kimura M and Tanaka H
  2006 {\em Phys. Rev. A\/} {\bf 74} 042704 publisher: American Physical
  Society \urlprefix\url{https://link.aps.org/doi/10.1103/PhysRevA.74.042704}

\bibitem{melton_radiolysis_1967}
Melton C~E and Rudolph P~S 1967 {\em J. Chem. Phys.\/} {\bf 47} 1771--1774 ISSN
  0021-9606 publisher: American Institute of Physics
  \urlprefix\url{https://aip.scitation.org/doi/abs/10.1063/1.1712163}

\bibitem{Sasic2004}
{\v{S}}a{\v{s}}i{\'{c}} O, Malovi{\'{c}} G, Strini{\'{c}} A, Nikitovi{\'{c}} Z
  and Petrovi{\'{c}} Z~L 2004 {\em New Journal of Physics\/} {\bf 6} 1--11 ISSN
  13672630

\bibitem{winters_dissociation_1975}
Winters H~F 1975 {\em J. Chem. Phys.\/} {\bf 63} 3462--3466 ISSN 0021-9606
  publisher: American Institute of Physics
  \urlprefix\url{https://aip.scitation.org/doi/10.1063/1.431783}

\bibitem{shirai_analytic_2002}
Shirai T, Tabata T, Tawara H and Itikawa Y 2002 {\em Atomic Data and Nuclear
  Data Tables\/} {\bf 80} 147--204 ISSN 0092-640X
  \urlprefix\url{https://www.sciencedirect.com/science/article/pii/S0092640X01908782}

\bibitem{salabas_two-dimensional_2002}
Salabas A, Gousset G and Alves L~L 2002 {\em Plasma Sources Sci. Technol.\/}
  {\bf 11} 448--465 ISSN 0963-0252 publisher: IOP Publishing
  \urlprefix\url{https://doi.org/10.1088/0963-0252/11/4/312}

\bibitem{frost_rotational_1962}
Frost L~S and Phelps A~V 1962 {\em Phys. Rev.\/} {\bf 127} 1621--1633
  publisher: American Physical Society
  \urlprefix\url{https://link.aps.org/doi/10.1103/PhysRev.127.1621}

\bibitem{al-amin_electron_1985}
Al-Amin S~A~J, Kucukarpaci H~N and Lucas J 1985 {\em J. Phys. D: Appl. Phys.\/}
  {\bf 18} 1781--1794 ISSN 0022-3727 publisher: IOP Publishing
  \urlprefix\url{https://doi.org/10.1088/0022-3727/18/9/009}

\bibitem{cochran_diffusion_1962}
Cochran L~W and Forester D~W 1962 {\em Phys. Rev.\/} {\bf 126} 1785--1788
  publisher: American Physical Society
  \urlprefix\url{https://link.aps.org/doi/10.1103/PhysRev.126.1785}

\bibitem{cottrell_drift_1965}
Cottrell T~L and Walker I~C 1965 {\em Trans. Faraday Soc.\/} {\bf 61}
  1585--1593 ISSN 0014-7672 publisher: The Royal Society of Chemistry
  \urlprefix\url{https://pubs.rsc.org/en/content/articlelanding/1965/tf/tf9656101585}

\bibitem{davies_measurements_1989}
Davies D~K, Kline L~E and Bies W~E 1989 {\em Journal of Applied Physics\/} {\bf
  65} 3311--3323 ISSN 0021-8979 publisher: American Institute of Physics
  \urlprefix\url{https://aip.scitation.org/doi/10.1063/1.342642}

\bibitem{heylen_ionization_1975}
Heylen A~E~D 1975 {\em International Journal of Electronics\/} {\bf 39}
  653--660 ISSN 0020-7217 publisher: Taylor \& Francis
  \urlprefix\url{https://doi.org/10.1080/00207217508920532}

\bibitem{hunter_electron_1986}
Hunter S~R, Carter J~G and Christophorou L~G 1986 {\em Journal of Applied
  Physics\/} {\bf 60} 24--35 ISSN 0021-8979 publisher: American Institute of
  Physics \urlprefix\url{https://aip.scitation.org/doi/10.1063/1.337690}

\bibitem{lakshminarasimha_ratio_1977}
Lakshminarasimha C~S and Lucas J 1977 {\em J. Phys. D: Appl. Phys.\/} {\bf 10}
  313--321 ISSN 0022-3727 publisher: IOP Publishing
  \urlprefix\url{https://doi.org/10.1088/0022-3727/10/3/011}

\bibitem{lin_moment_1979}
Lin S~L, Robson R~E and Mason E~A 1979 {\em J. Chem. Phys.\/} {\bf 71}
  3483--3498 ISSN 0021-9606 publisher: American Institute of Physics
  \urlprefix\url{https://aip.scitation.org/doi/10.1063/1.438738}

\bibitem{millican_electron_1987}
Millican P~G and Walker I~C 1987 {\em J. Phys. D: Appl. Phys.\/} {\bf 20}
  193--196 ISSN 0022-3727 publisher: IOP Publishing
  \urlprefix\url{https://doi.org/10.1088/0022-3727/20/2/007}

\bibitem{pollock_momentum_1968}
Pollock W~J 1968 {\em Trans. Faraday Soc.\/} {\bf 64} 2919--2926 ISSN 0014-7672
  publisher: The Royal Society of Chemistry
  \urlprefix\url{https://pubs.rsc.org/en/content/articlelanding/1968/tf/tf9686402919}

\bibitem{shimozuma_measurement_1981}
Shimozuma M and Tagashira H 1981 {\em J. Phys. D: Appl. Phys.\/} {\bf 14}
  1783--1789 ISSN 0022-3727 publisher: IOP Publishing
  \urlprefix\url{https://doi.org/10.1088/0022-3727/14/10/012}

\bibitem{particle_swarm}
{P}article {S}warm {B}oltzmann {S}olver
  \urlprefix\url{www.gitlab.com/MD-CWI-NL/particle_swarm}

\bibitem{Teunissen2016}
Teunissen J and Ebert U 2016 {\em Plasma Sources Science and Technology\/} {\bf
  25} ISSN 13616595

\bibitem{Sebastian2005}
Sebastian A~A and Wadehra J~M 2005 {\em Journal of Physics D: Applied
  Physics\/} {\bf 38} 1577--1587 ISSN 00223727

\bibitem{brigg_r-matrix_2014}
Brigg W~J, Tennyson J and Plummer M 2014 {\em J. Phys. B: At. Mol. Opt.
  Phys.\/} {\bf 47} 185203 ISSN 0953-4075 publisher: IOP Publishing
  \urlprefix\url{https://doi.org/10.1088/0953-4075/47/18/185203}

\bibitem{habershon_ring-polymer_2013}
Habershon S, Manolopoulos D~E, Markland T~E and Miller T~F 2013 {\em Annu. Rev.
  Phys. Chem.\/} {\bf 64} 387--413 ISSN 0066-426X publisher: Annual Reviews
  \urlprefix\url{https://www.annualreviews.org/doi/10.1146/annurev-physchem-040412-110122}

\bibitem{LxCat}
{LXC}at \urlprefix\url{www.lxcat.net}

\bibitem{perrin_dissociation_1982}
Perrin J, Schmitt J~P~M, de~Rosny G, Drevillon B, Huc J and Lloret A 1982 {\em
  Chemical Physics\/} {\bf 73} 383--394 ISSN 0301-0104
  \urlprefix\url{https://www.sciencedirect.com/science/article/pii/030101048285177X}

\bibitem{Rohatgi2020}
Rohatgi A 2020 Webplotdigitizer: Version 4.4
  \urlprefix\url{https://automeris.io/WebPlotDigitizer}

\end{thebibliography}
\end{document}